\documentclass[10pt,journal,compsoc]{IEEEtran}
\usepackage{makecell}
\usepackage{amsmath}
\usepackage{algorithm}
\usepackage[noend]{algpseudocode}
\usepackage{enumitem}
\usepackage{graphicx}
\usepackage{multirow}
\usepackage{lscape}
\usepackage{mathtools}
\usepackage{gensymb}
\usepackage{booktabs}
\usepackage{ragged2e}

\usepackage{todonotes}

\ifCLASSOPTIONcompsoc
  \usepackage[nocompress]{cite}
\else
  \usepackage{cite}
\fi

\begin{document}
%
\title{
Drone-as-a-Service Composition Under Uncertainty
}

\author{Ali~Hamdi, Flora~D.~Salim, Du~Yong~Kim, Azadeh~Ghari~Neiat, and~Athman~Bouguettaya
\thanks{Ali Hamdi and Flora D. Salim are with the School of Computing Technologies, RMIT University, Melbourne, Australia. Emails: \{ali.ali, flora.salim\}@rmit.edu.au.}
\thanks{Du Yong Kim is with the School of Engineering,  RMIT University, Melbourne, Australia, Email: duyong.kim@rmit.edu.au.}

\thanks{Azadeh~Ghari~Neiat is with the School of Information Technology, Deakin University, Geelong, Australia,  Email:azadeh.gharineiat@deakin.edu.au}
\thanks{Athman~Bouguettaya is with the School of Computer Science,  University of Sydney, Sydney, Australia. Email: \{azadeh.gharineiat, athman.bouguettaya\}@sydney.edu.au}
}

\markboth{IEEE Transactions on Services Computing}%
{Hamdi \MakeLowercase{\textit{et al.}}: Uncertainty-aware Drone-as-a-Service Scheduling, Route-Planning, and Composition}

\IEEEtitleabstractindextext{%
\begin{abstract}
\justifying
We propose an uncertainty-aware service approach to provide drone-based delivery services called Drone-as-a-Service (DaaS) effectively. Specifically, we propose a service model of DaaS based on the dynamic spatiotemporal features of drones and their in-flight contexts. 
The proposed DaaS service approach consists of three components: scheduling, route-planning, and composition. 
First, we develop a DaaS scheduling model to generate DaaS itineraries through a Skyway network. 
Second, we propose an \emph{uncertainty-aware DaaS route-planning algorithm} that selects the optimal Skyways under weather uncertainties.
Third, we develop two DaaS composition techniques to select an optimal DaaS composition at each station of the planned route.  
A \emph{spatiotemporal DaaS composer} first selects the optimal DaaSs based on their spatiotemporal availability and drone capabilities. A \emph{predictive DaaS composer} then utilises the outcome of the first composer to enable fast and accurate DaaS composition using several Machine Learning classification methods. We train the classifiers using a new set of spatiotemporal features which are in addition to other DaaS QoS properties. Our experiments results show the effectiveness and efficiency of the proposed approach.
\end{abstract}

\begin{IEEEkeywords}
Drone-as-a-Service, uncertainty-aware, service scheduling, route-planning, composition. 
\end{IEEEkeywords}}

\maketitle
\IEEEdisplaynontitleabstractindextext
\IEEEpeerreviewmaketitle

\IEEEraisesectionheading{\section{Introduction}
\label{sec:introduction}}
\IEEEPARstart{T}{he} commercial drone market is expected to reach \$127 billion in 2020 \cite{pwc2016}. Pervasive Internet-of-Things (IoT) embedded sensors provide new drones usages to improve human lifestyle and productivity. 
New research opportunities are opened in different computing fields such drone learning \cite{hambling2015drone}, IoT \cite{koubaa2017service}, computer vision \cite{hamdi2020drotrack}, and smart environments \cite{loke2016smart}. 
This leads to a wide variety of new applications ranging from public security \cite{He-Drone}, 
disaster management \cite{luo2019unmanned}
, E-commerce \cite{Krakowczyk-Developing, Kim-Cost}
, and delivery \cite{Dorling-Vehicle,Krakowczyk-Developing}. Using drones in delivery services is a fast-growing industry that gained much attention from companies such as Amazon, DHL, and Google. Drones offer low-cost delivery services in comparison to conventional methods. As reported in  \cite{Keeney2015Amazon}, the cost of small parcel delivery for a 16.1 km distance in the USA is only \$1 using a drone while it costs \$12.92 and \$ 8.32 using UPS and FedEx delivery services.

The \emph{service} paradigm is leveraged as a pivotal mechanism to model the drone-based delivery as a service called Drone-as-a-Service (DaaS). DaaS is the abstraction of the drone delivery functionalities along with their qualities, a.k.a., Quality of Services (QoS) properties of drone services. This abstraction will facilitate the implementation of large-scale delivery strategies despite the inherent challenges imposed by drones such as limited flight range and battery capacity. For instance, a single short-range drone may not guarantee the delivery of a package to a long-distant destination. Composing multiple DaaS enables the combination of different drones to reach distant delivery destinations. DaaS composition aims at maximising the drone performance and minimising the operational costs \cite{park2016design}. 

There is a large body of related works that focused on DaaS connectivity between the drone and the ground station \cite{Zeng2018Network}, and DaaS resources limitations \cite{Yanmaz2017Communication}. However, there has been little attention in existing research which address the DaaS delivery related research challenges, e.g., \textit{uncertainty}. Drone delivery services present more complex challenges as they require both \textit{cyber} (e.g., route-planning), and \textit{physical} (e.g., delivery pickup and drop-off), and processes. 
Because of the nature of drones, uncertainty is an intrinsic part of the drone delivery service environment. \emph{Uncertainty} may be caused by drone capabilities, payload limitations, and real-time variation of environmental aspects including precipitation (e.g., rain, snow, and hail), wind (e.g., power, direction, and altitude), temperature, and visibility  \cite{Kim-18}.   
Weather uncertainty is a crucial challenge; however, uncertainty can also be found in other aspects such as package weights, battery levels and lifetime. This uncertainty introduces challenges of qualitative factors, such as availability. As a result, there is a need for an uncertainty-aware drone service selection and composition algorithms. 

We propose and develop an uncertainty-aware spatiotemporal DaaS approach. The proposed approach considers the uncertainty effects of weather, drone capabilities, spatiotemporal availability, and computational costs on DaaS selection and composition.
The proposed approach consists of three components: \textit{DaaS scheduling}, \textit{DaaS route-planing}, and \textit{DaaS composition}.
In the first component, we propose a new DaaS scheduling model that considers both cyber and physical processes. The proposed model constructs a delivery network that consists of delivery stations and Skyways. We assume that all the DaaS requests are known in advance. We design the use case to have fixed delivery stations and DaaS requests.
In the second component, we propose \emph{uncertainty-aware route-planning} that predicts spatiotemporal availability of a drone service considering weather uncertainty, drone performance and delivery's payloads. In the third component, we propose a \emph{spatiotemporal DaaS composer} for the optimal spatiotemporal composition of drone delivery services to meet desired functional and QoS requirements under a range of uncertainties. However, a key challenge is an expensive computation that is required to handle these different issues. We propose a \emph{predictive DaaS composer} to overcome the expected latency due to the required high-computational costs. The predictive composer trains classification models to achieve accurate, high-speed DaaS service compositions. It also enables more computationally-efficient DaaS composition. However, the accuracy of the uncertainty-aware DaaS composer depends on the quality of the collected data, such as weather data. Therefore, we deal with the problem of uncertainty that exists in the situation inputs of the DaaS delivery services such as weather measurements.

Our main contributions are summarised as follows:
\begin{itemize}[leftmargin=1em]
\item A novel architecture for uncertainty-aware DaaS design and scheduling. The proposed design considers the best route and optimal drones at different locations, times, and weather conditions to generate service schedules.
\item An uncertainty-aware route-planning algorithm based on  A* \cite{Hart1968A}. The proposed algorithm defines a new heuristic function which uses the weather conditions and drone capabilities to calculate the relative drone airspeed, distance, and flying duration. The algorithm solves the issue of weather uncertainty to select the best Skyways.
\item A spatiotemporal DaaS composer includes spatial, temporal, and weight-aware DaaS composition layers that offer optimal DaaS service composition plan.
\item A predictive DaaS composer produces accurate, high-speed DaaS-composition. This algorithm utilises the DaaS composition outcomes of the spatiotemporal algorithm to train classification models. It aims to produce real-time DaaS compositions.
\end{itemize}

{The rest of the paper is organised as follows. Section \ref{related-works} reviews the related work in the area of drone-based delivery, context-aware, drone route planning and spatio-temporal service composition. We then describe the motivation scenario in Section \ref{Motivation} and the DaaS service design in Section \ref{DaaS-service-design}. The approach components are discussed in multiple sections, including, scheduling in Section \ref{scheduling}, route-planning in Section \ref{route-planning}, spatiotemporal composer in Section \ref{STcomposer}, and predictive composer in Section \ref{Pcomposer}. The experimental results are discussed in Section \ref{ExpRes}. Section \ref{conclusion} concludes the paper.}
\begin{table}[]
\centering
\caption{A list of the used abbreviations and their explanations.}
\label{tbl:abb}
\begin{tabular}{|p{1cm}|p{2.5cm}|p{1cm}|p{2.5cm}|}
\hline
Abbre-viation & Explanation                 & Abbre-viation & Explanation                                      \\ \hline
DaaS         & Drone-as-a-Service          & DSP          & Delivery service provider                        \\ \hline
PDR          & Package delivery requests   & QoS          & Non-functional or Quality of Services properties \\ \hline
$P_{loc}$    & Pickup location             & $P_{t}$      & Pickup time                                      \\ \hline
$D_{loc}$    & Delivery location           & $D_{t}$      & Delivery time                                    \\ \hline
Dist.        & Distance                    & fd           & Flight duration in minutes                       \\ \hline
pl           & payload capacity            & sp           & Maximum flying speed                             \\ \hline
WS           & Wind speed                  & WB           & wind bearing                                     \\ \hline
T            & Temperature                 & Hu           & Humidity                                         \\ \hline
V            & Visibility                  & CC           & Cloud cover                                      \\ \hline
DP           & Dew point                   & DAS          & Drone air speed                                  \\ \hline
GSA          & Ground speed average        & CM           & Certainty Margin                                 \\ \hline
SGD          & Stochastic Gradient Descent & SVM          & Support Vector Machine                           \\ \hline
LR           & Logistic Regression         & GNB          & Gaussian Naive Bayes                             \\ \hline
\end{tabular}
\end{table}
\section{Related Work}\label{related-works}
We overview the related research in drone-based delivery, situation-aware service composition, drone route-planning and spatiotemporal service selection and composition.

There is a body of related works that focused on the drone-based delivery \cite{Dorling-Vehicle, Krakowczyk-Developing, park2016design}.
\cite{park2016design} proposed a service selection model for large-scale delivery services. The proposed approach depends on the energy consumption optimisation to maximise profitability and minimise the service delivery time. Moreover, \cite{ferrandez2016optimization} investigated the integration of truck-drone for delivery services. Their solution estimates the optimal number of launch locations utilising k-means clustering and determines the best delivery route using a genetic algorithm. However, these approaches do not consider the weather impact on the delivery cost, time or route. \cite{gatteschi2015new} implemented a system for drug-delivery services. Their system focused on the delivery precision, and drone preservation. It uses the weather information for the distance area between the service provider and the consumer locations. However, the proposed framework asks the service provider to decide whether to fly the drone or to delay instead of computing the best route based on the retrieved weather information. 

The IoT and service computing researchers have developed context-aware service selection, and composition algorithms \cite{li2011semantic, Fissaa2018, Badidi2018}. An algorithm for context-aware service composition was developed in \cite{li2011semantic} to compose the Web services as the context changes. 
In \cite{yu2012multi}, a multi-attribute optimisation approach for service selection is proposed. They presented a set of service composition techniques. These techniques select the optimal services that match requirements and include the user desired providers. The authors in \cite{Fissaa2018} proposed an approach for context-aware service selection. They employed Markov Decision process with reinforcement learning techniques.
The authors in \cite{Badidi2018} proposed a service adaptation and personalization framework. The framework considers the user profile, the user context, and the composition specification. They use the user's mobile device and external services to collect the context information. 
The quality of the context guides the service selection. However, the delivery services are more complicated as they require both cyber and physical processes. Therefore, it is necessary to develop situation-aware service selection and composition algorithms. 

Finding the optimal route is a crucial process in various applications such as parking monitoring \cite{shao2018traveling,qin2020solving}, crime monitoring and emergency management \cite{rumi2020realtime}, and drone service delivery \cite{chang2018optimal}. The work in \cite{Kumar-MVO-Based} used multiverse optimiser (MVO) to solve the drone path planning problem. 
\cite{yongbo2017three,zhang2016grey} developed drone path planning using a modified pack search and grey wolf optimiser algorithms. Other techniques were developed based on vision algorithms \cite{phung2017enhanced}. Although there exists research on the drone route planning as mentioned, a very few approaches focus on dynamic route planning for drone delivery services. 

There exist studies to select IoT services based on spatiotemporal features \cite{chang2018optimal,Lakhdari-Crowdsourcing, Kumar-MVO-Based}. For example, a spatiotemporal selection and composition framework for sensor-cloud services was proposed in \cite{neiat2014spatio,neiat2014failure}. Spatiotemporal features are the focal aspects of the service model. In the model, a service consists of several functional attributes and associated QoS. In particular, new spatiotemporal QoS attributes are proposed to evaluate sensor-cloud services based on the spatiotemporal properties of the services. A spatiotemporal linear composition algorithm is developed, which enables users to select their desired sensor-cloud services based on multiple criteria. 
Uncertainty is an additional complicating factor for drone service composition and selection. Uncertainty is due to the spatiotemporal dynamicity and real-time variation of environmental factors. For example, QoS attributes of a drone service such as flight range and state of the battery are dependent on the weather conditions and weight of payloads. There are very few approaches that focus on dynamic selection and composition of drone service considering uncertain QoS parameters. 
In this paper, we propose to address this research gap by developing a novel uncertainty-aware DaaS service composition framework. We focus on finding the optimal DaaS composite services based on the available drones capabilities and weather conditions. 

\section{Uncertainty-aware DaaS Motivation Scenario}
In this section, we explain the problem scenario and the proposed DaaS delivery service model that considers the motivating scenario constraints. Table \ref{tbl:abb} lists and explains the abbreviations that are used in the rest of this paper.

\begin{figure}[h!]
\centering
\includegraphics[width=0.48\textwidth]{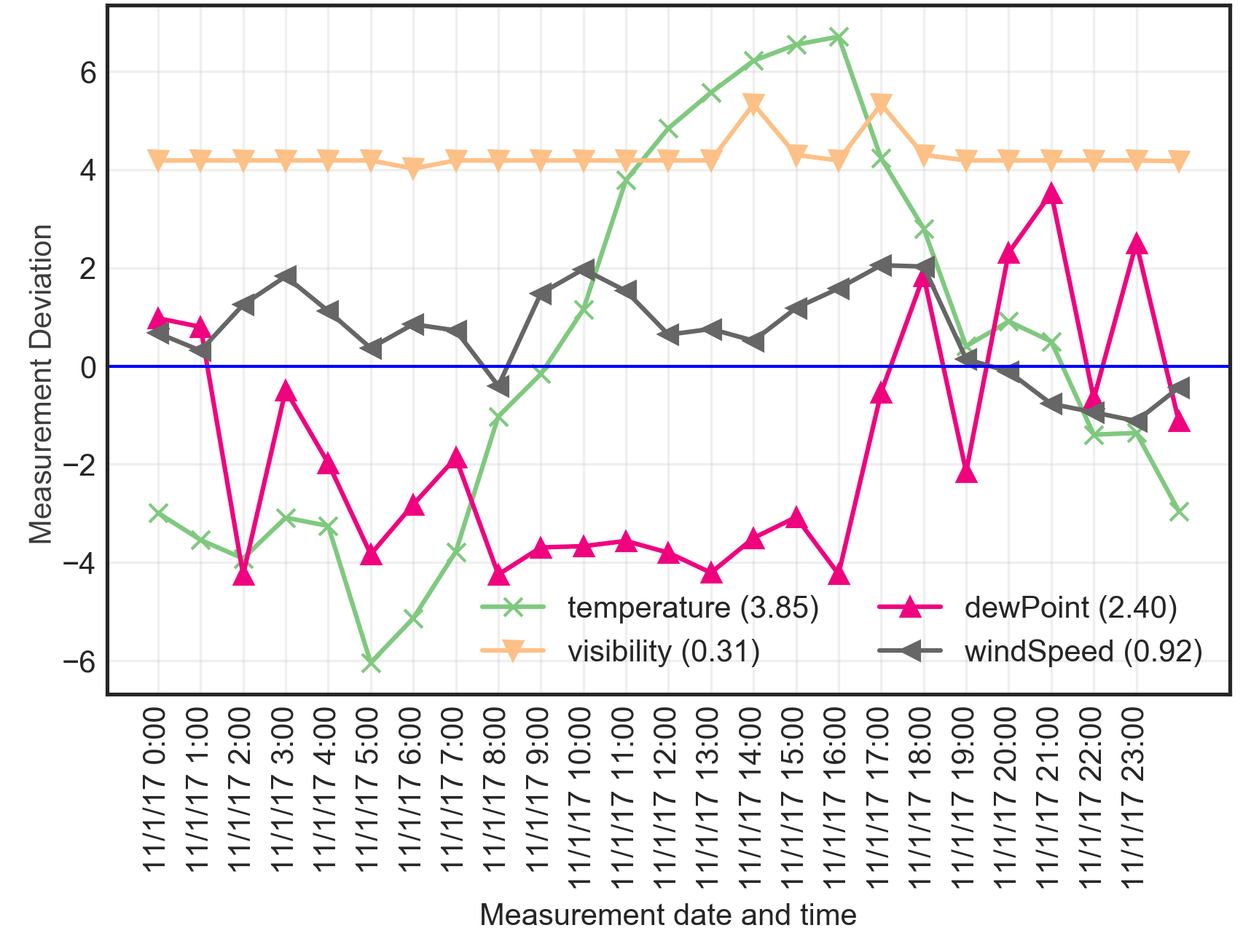}
\caption{An uncertainty analysis highlights the weather measurements deviations across 24 hours. The figure shows each weather attribute accompanied by its measurement standard deviation.}
\label{24_errors}
\end{figure}

\subsection{Problem Definition}\label{Motivation}
In many commercial applications, drones can be used to carry packages such as books, drugs or fast food. We assume that a delivery service provider (DSP) has a fleet of drones. These drones have different battery levels, flight duration, endurance, payloads, and speeds. The DSP aims to carry delivery packages among a set of fixed delivery stations as designed in Fig. \ref{DaaS_network_graph}. We assume that there is a Skyway network within a specific geographic region that connects stations. We also assume that the Skyway network is divided into predefined Skyways. The Skyway network is assumed to follow the regulation rules to avoid no-fly zones and controlled air-spaces. Each drone has a predefined itinerary to fly through a certain path, i.e., one or more Skyways, in the network. The DSP receives a set of package delivery requests (PDR). Each PDR has a weight, pickup and drop stations (locations) and delivery times. A single drone may not be able to travel for long delivery distances without recharging. In such a case, multiple drone services may be combined into a composite drone. 

We reformulate the problem of drone delivery service composition as finding an optimal set of drone delivery services from a warehouse station to a delivery station. To this end, the selection and composition of DaaS services are proposed to provide the best quality of delivery service avoiding bad weather condition, following the regulations, maximising battery lifetime, and minimising the delivery time.
Fig. \ref{24_errors} shows an uncertainty analysis of multiple weather measurements. The blue-coloured horizontal line represents 0 deviations with the actual weather measurements. Fig. \ref{24_errors} illustrates that the uncertainty problem is affecting most of the weather measurements. For example, visibility forecast measurements are always higher than the actual measurements. We focus on the weather uncertainty; however, it can be found in other aspects such as uncertainty of package weights, battery levels and lifetime.


In this paper, we focus on the scenario of the delivery across fixed delivery stations and predefined sky-paths. This scenario can be applied for e-commerce or governmental organisation.
It is useful for a situation such as when a constructing company moves building materials and equipment among different work locations. They would have a preliminary plan of what to more and when. The plan can be considered to have multiple delivery requests in a sequence. This can also be applied in other examples such as mining companies, defence and border protections operations. Such domains may require moving reassures among different locations according to different planned schedules. The proposed framework can also be deployed in other use-cases such as lifesaving delivery \footnote{flyzipline.com}.

{In the following sections, we explain the new DaaS model and how it is employed to achieve an uncertainty-aware DaaS composition.}
%
%

\subsection{DaaS Service Model}\label{DaaS-service-design}

We define a new service model for a DaaS that flies through Skyways between the pickup and drop-off locations during a period starts at pickup time and ends at drop-off time. The DaaS is modelled as follows. 

\textit{Definition 1: DaaS Service Model.} $DaaS\ is\ a\ tuple\ of\ < ID, F, QoS,  P_t, P_{loc}, D_t, D_{loc}>$ where: 

\begin{itemize}
\item \emph{ID} is a a unique identifier for a drone service, 
\item \emph{F} describes a set of drone functions (e.g., delivery or sensing), 
\item \emph{QoS} is a tuple $<$ $q_1,q_2, ..., q_n$ $>$, where each $q_i$ denotes a QoS property of DaaS, 
\item \emph{$P_t$} is the \emph{pickup time} or a scheduled departure time at which the DaaS will pick up the package and fly,
\item \emph{$P_{loc}$} represents the \emph{pickup location} where the DaaS pick up the delivery package from,
\item \emph{$D_t$} is the \emph{drop-off time} or a scheduled arrival time for the DaaS to drop the delivery package, and
\item \emph{$D_{loc}$} is the \emph{drop-off location} where the DaaS will deliver the package.
\end{itemize}
\textit{Definition 2: DaaS QoS Model.}
We propose a QoS model that introduces new QoS attributes for DaaS which focus on dynamic aspects of drone as follows.
\begin{equation}\label{QoS}
QoS \Rightarrow < fd, pl, sp, Env> 
\end{equation}
where; 
\begin{itemize}
\item \emph{fd} is the drone initial \emph{flight duration} (minutes) during good weather conditions. It is calculated based on the DaaS scheduled departure and arrival time, 
\begin{equation}
    fd = \frac{Flight Distance (km)}{ Drone Speed (km/h)} * 60
\end{equation}
\item \emph{pl} is the drone  maximum \emph{payload capacity}. The composition algorithm uses it to select the DaaS to achieve weight-aware selection. In which the \emph{pl} will be compared to the PDR weight.,
\item \emph{sp} is the drone maximum flying \emph{speed}. The composition algorithm uses \emph{sp} as one of the heuristics to find the optimal path.
\end{itemize}

\textit{Definition 2: Package Delivery Request.}
A Package Delivery Request (PDR) is defined as a service request to deliver a package from a pickup location to drop-off location.
$PDR\ is\ defined\ as\ a\ tuple\ of < P_{loc}, P_t, D_{loc}, w, rt >$ where;

\begin{itemize}
\item \emph{$P_{loc}$} \& \emph{$P_{t}$} is the pickup location and time, respectively,
\item \emph{$D_{loc}$} is the drop-off location, 
\item \emph{w} is the \emph{weight} of the PDR, and
\item \emph{rt} is the request time-stamp.
\end{itemize}

\section{Uncertainty-aware DaaS Approach}
\begin{figure*}[h!]
\centering
\includegraphics[width=.8\textwidth]{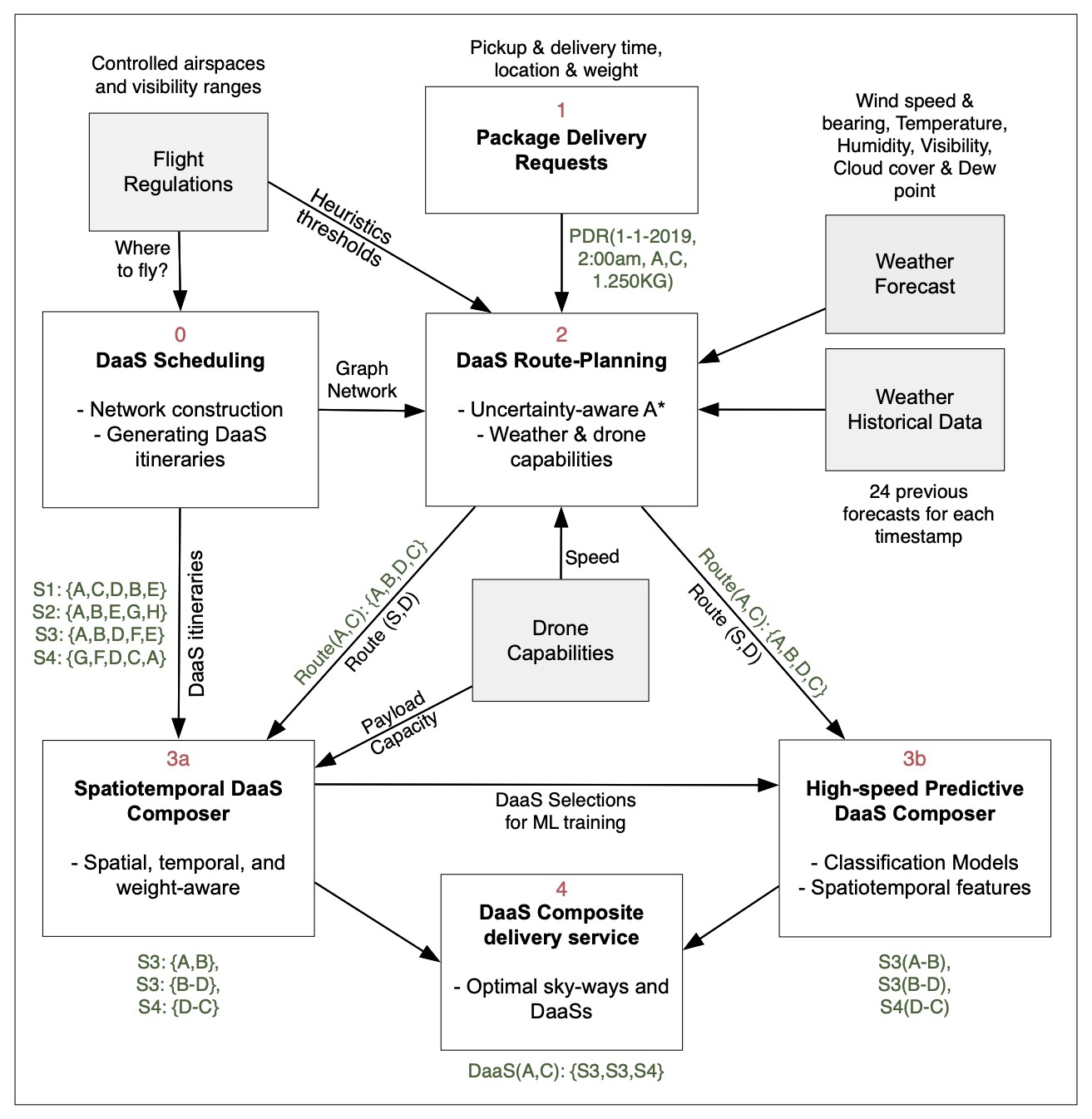}
\caption{The figure connects the main components of the proposed uncertainty-aware DaaS approach with the employed datasets. Component 1 represents the process of receiving the PDR and passing it to the route-planning at 2. The best route is sent to the spatiotemporal DaaS composer, numbered with 3a, which utilise the itineraries from 0 to produce the optimal DaaS composite service at 4. The selections of 3a are passed to 3b to train classification models. These models are used after 1, 2 to produce the composite DaaS at 4.}
\label{DaaS-approach}
\end{figure*}

This section explains the three main components of the proposed \textit{uncertainty-aware DaaS approach}: DaaS scheduling, route-planning, and composition. Fig. \ref{DaaS-approach} shows these components and utilised datasets and information. The first component consists of the DaaS scheduling model. The model is labelled with 0 as it is an offline process to construct the Skyway network graph and generate the drones itineraries (more details in Section \ref{scheduling}). In the second component, an \emph{uncertainty-aware route-planning} algorithm is proposed to select the best route of Skyways in the first component (see Section \ref{route-planning}). The composition component iteratively finds an optimal DaaS service at each delivery stations of the best route. We propose two DaaS composition algorithms to find an optimal DaaS composition plan for the incoming PDR: \emph{spatiotemporal DaaS composer} which considers the spatiotemporal and drone payload capacity and \emph{predictive DaaS composer} to produce fast, low-computation compositions (see Sections \ref{STcomposer} and \ref{Pcomposer} for more details). {Fig. \ref{DaaS-approach} shows an example, green-coloured texts, that receives a PDR to deliver a 1.250 kg from station A to C at 2:00 am, 11/1/19.} The PDR details are passed to the DaaS uncertainty-aware route-planning algorithm. The algorithm computes the best route through the network stations A, B, D, and C. Then, the spatiotemporal DaaS composer receives the route nodes and available DaaS scheduled itineraries, e.g., S1, S2, S3, and S4, to select the optimal DaaS. The outcome selections are combined to represent the final composite DaaS, see label 4. Those selections are also being used to train machine learning classification models. In the case that the selections are enough to train accurate models, the DaaS selections will be made as predictions at each node of the best route.  
\section{DaaS Scheduling Model}\label{scheduling}
We propose a DaaS service scheduling model that contains three main components, including, 
1) Skyways network construction, 
2) DaaS delivery services preparation and itineraries scheduling, and
3) DaaS simulation and data collection. The outcome of this model is used by the proposed DaaS route-planning and composition algorithms.

\subsection{Skyways Network Construction}
\begin{figure}[h!]
\centering
\includegraphics[width=0.42\textwidth, angle=0,origin=c]{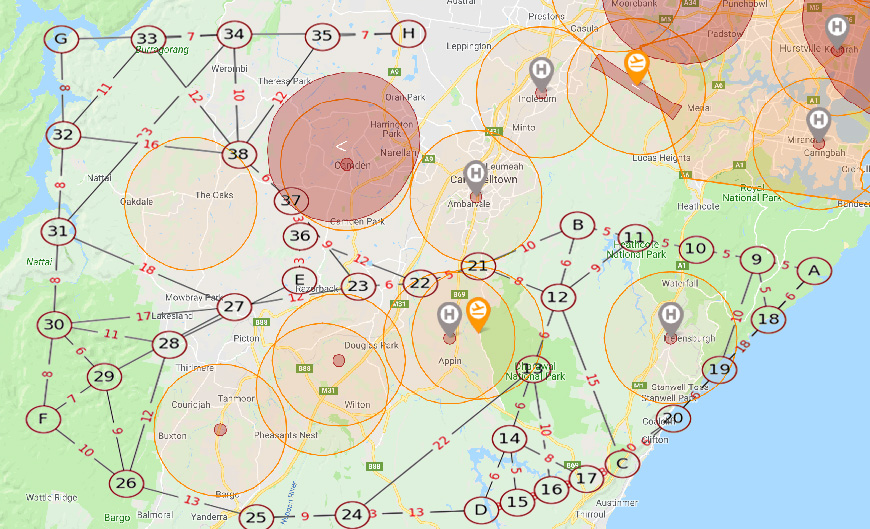}
\caption{DaaS network graph with distances of edges in km.}
\label{DaaS_network_graph}
\end{figure}

{According to our delivery scenario, we construct a Skyways network that consists of $1,254$ delivery stations that are connected by $1,280$ Skyways.
For simplicity, we discuss the uncertainty issues and DaaS composition on a small network of $38$ fixed delivery stations and $64$ Skyways.} 
A Skyway is defined as a line segment between two particular delivery stations. Table \ref{network-Skyways} (Appendix \ref{Appendix_data}) lists some examples of the Skyways data. For example, the Skyway \emph{F-DS\_30} is the line segment that connects the delivery stations \emph{F} and \emph{DS\_30}. The table also lists the distances in kilometres and the compass bearing, i.e., the heading direction, for each Skyway. Fig. \ref{DaaS_network_graph} shows the constructed network graph. The nodes of the graph represent the delivery stations. For better visualisation, we removed the 'DS\_' from the stations' names at the nodes. The red numbers at the edges among the delivery stations are the distances in km.
Fig. \ref{DaaS_network_graph} shows the regulations which affect the design of the Skyways network. The map in Fig. \ref{DaaS_network_graph} is provided by an online service that is operated by The Civil Aviation Safety Authority (CASA) in Australia. 
It has two different types of coloured circles. The red-coloured circles refer to no-fly areas under controlled air-spaces. The orange-coloured circles represent warning areas such as within 5.5 kilometres of airports \footnote{For more details about CASA regulations: \emph{casa.gov.au}.}. Using Google Maps, we extracted the GPS coordinates for the stations and Skyways. Fig. \ref{DaaS_network_graph} shows the network graph.
%
%
%
%
%
%
\subsection{DaaS Delivery Services Scheduling}
We assume that the DSP has a fleet of two types of drones. We use two drone models manufactured by DJI \footnote{https://www.dji.com/.} with different capabilities. DJI Phantom 4 PRO carries up to 0.5 kg, and DJI Matrice 200 carries up to 2.34 kg. 
Table \ref{drone-types}
highlights the main features of each type. 
According to the motivating scenario, the DSP will fly the drones following predefined schedule itineraries. We assume that the delivery station \emph{A}, \emph{C}, \emph{E}, \emph{G}, and \emph{H} are major stations. Therefore, the source and destination of a DaaS delivery service have to be from those major stations.
Table \ref{DaaS-service} (Appendix \ref{Appendix_data}) lists some examples of DaaS delivery services. Each service is represented spatially to start from a source delivery station, lands, in between, at certain stations, stops at a final destination. Table \ref{DaaS-service} (Appendix \ref{Appendix_data}) also shows the timetable for each service. We assume that the DSP will fly many services through each path. However, each service has a different pickup time, source, destination, drop-off time, and a certain type of drone. We assume that a drone requires a maintenance time duration, i.e., a period required for battery changing and maintenance checkup. The battery replacement technique is known in the literature as battery hot-swapping \cite{lee2015autonomous} in which the drone can automatically land in a station and replace its battery within 60 seconds. Moreover, the flight duration \emph{fd} is expected to be more than the planned duration due to the acceleration and deceleration. The maintenance time tends to cover such uncertainty. For experimental work, we use 5 and 15 minutes as maintenance time durations for the Phantom 4 PRO and Matrice 200, respectively. We estimate the \emph{fd} and drop-time based on the given service flight distance, pickup time, maintenance duration, and drone speed. 
{The \emph{fd} in Table \ref{DaaS-service} (Appendix \ref{Appendix_data}) are calculated regardless of the weather uncertainty for planning purposes. Later, the proposed DaaS composition algorithm will consider the weather impact on such measurements. }

\begin{table}[h!]
\centering
\caption{The main capabilities for two models of DJI multi-rotors drones.}
\label{drone-types}
\begin{tabular}{|p{2.5cm}|p{2.5cm}|p{2.5cm}|}
\hline
Feature & DJI PHANTOM 4 PRO & DJI MATRICE 200 \\ \hline
Drone Type & multirotor-4 & multirotor-4 \\ \hline
Flight time & 30 mins & 38 mins \\ \hline
Speed & 72 km/h & 82.8 km/h \\ \hline
Base Weight & 1.388 kg & 3.80 kg \\ \hline
Maximum Takeoff Weight & 1.888 kg & 6.14KG \\ \hline
Delivery Payload & 0.50 km & 2.34 kg \\ \hline
Battery capacity & 5870 mAh & 7660 mAh \\ \hline
Max Wind Speed Resistance & 29 - 38 km/h & 43.2 km/h \\ \hline
Operating Temperature Range & $0^\circ$C - $40^\circ$C & -$20^\circ$ to $45^\circ$ C \\ \hline
Maintenance Time & 5 mins & 15  mins \\ \hline
\end{tabular}
\end{table}

%
\subsection{DaaS Data Collection}\label{Data-Collection}
\begin{figure}[]
\centering
\includegraphics[width=0.48\textwidth]{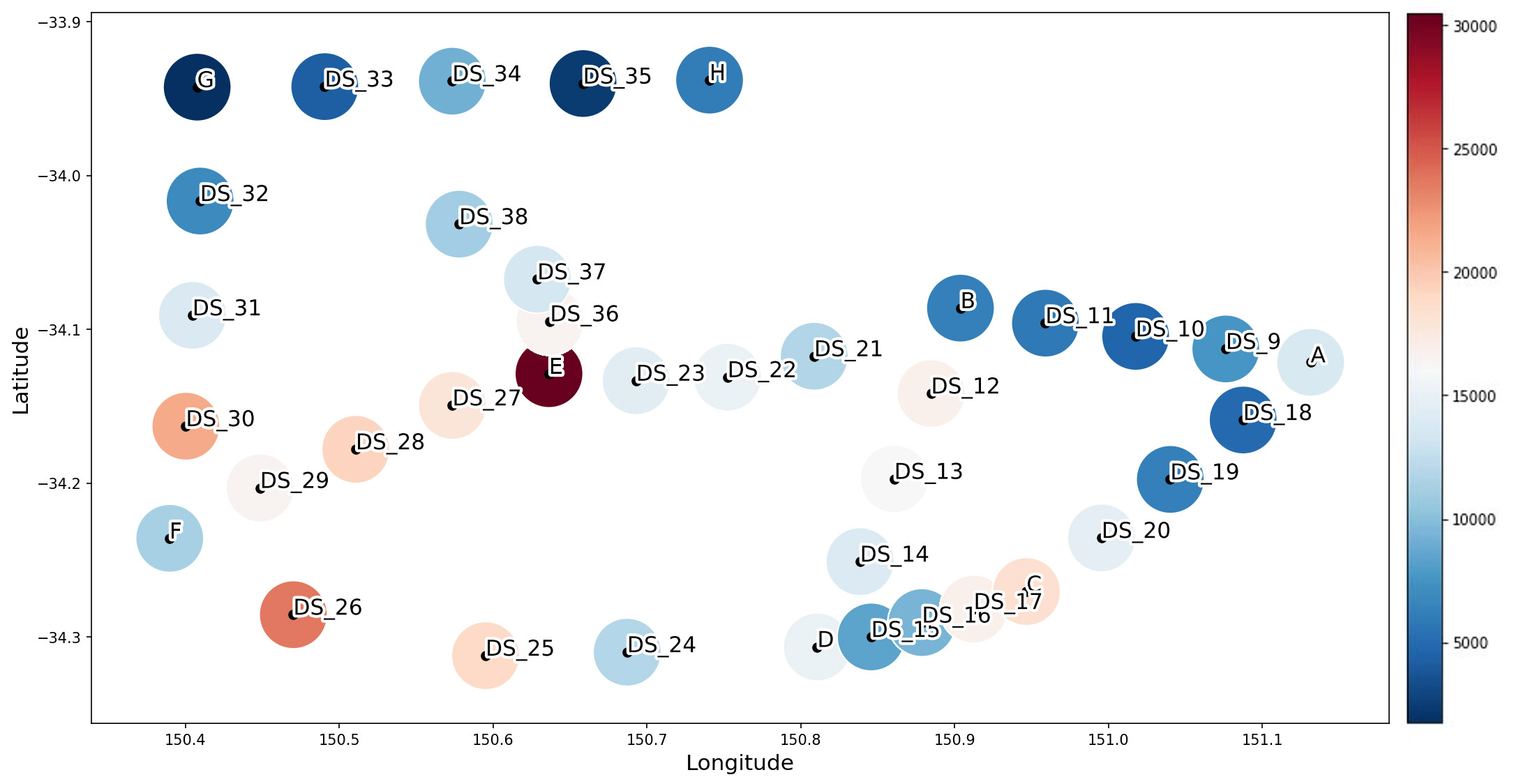}
\caption{The figure shows the closeness distributions of the DaaSs at each delivery station. The blue colour represents the less number of services and red colour represents the large numbers.}
\label{DaaS_distribtions}
\end{figure}

{
We generated small and large datasets using the above-mentioned DaaS delivery services scheduling and drone types. The large network has $1,254$ delivery stations as nodes, $1,280$ Skyways as edges, and the two drone types to define $16,770 $ DaaS delivery services. The small network has $38$ nodes and $64$ edges or Skyways with $30,476$ drone services. 
Table \ref{DaaS-service} shows a sample of the generated DaaS dataset. For example, the DaaS number 1 flies from the source station \textit{E} to the destination \textit{A}. The total distance of the service flying is $169.19$ km, consuming $197.6$ minutes of flying and maintenance over $16$ segments. The generated services are scheduled within two time-ranges. First, it starts from 1st to the 9th of November, 2017. Second, it starts from 1st to the 9th of May, 2018. We simulated the movements of the generated services between the delivery stations. Table \ref{DaaS-simulation} (Appendix \ref{Appendix_data}) lists the first movements of the DaaS with ID = 1. The total number of simulated movements is $4,589,290$ for the large network and $478,398$ for the small one. 
Fig. \ref{DaaS_distribtions} highlights the distributions of the simulated services across the different delivery stations. It shows the closeness density of DaaSs at each delivery station. The dark-red colour represents service counts greater than $30,000$, and the dark-blue is for less than $5,000$. For example, Fig. \ref{DaaS_distribtions} shows that the delivery station \textit{E} hosts the largest number of services. This is due to the central location of \textit{E} that enables connecting such a large number of services. We used the \emph{darksky.net} API to collect the forecast weather measurements at the source and destination coordinates for each Skyway. We collected $1,294,185$ and $16,000$ records of the weather data within the predefined two time-ranges for the two networks, i.e, large and small, respectively. The collected API data include various weather attributes such as temperature, cloud cover, wind speed, wind bearing, humidity, pressure, and dew-point. 
}
%
%
%
%

\section{DaaS Package Delivery Request (PDR)}
This section explains how to receive PDRs and prepare the system to process it, see label 1 in Fig. \ref{DaaS-approach}. Algorithm \ref{DaaS_Composer} shows the pseudo-code of receiving the PDR, compute the heuristics, find the best route, and the DaaS composition. The algorithm starts with receiving a PDR. The PDR requires to pick up a delivery package from a source to a destination station. It includes package weight, pickup time, and request timestamp. This PDR component involves four preliminary functions. It consists of (1) sorting the PDR request, (2) temporal domain definition, (3) station and Skyways geo-locating, and (4) weather dataset filtration.
\emph{DaaS PDR sorting} is a process to sort incoming PDR requests according to their query timestamps \emph{rt}. This sorting process aims to prioritise the PDR request based on the request time order.
\emph{Temporal domain definition} is a process to define a time window duration, e.g., 15 minutes, to expand the PDR pickup time. Instead of only selecting from DaaSs that fly on the same \emph{rt}, the temporal domain will act to aggregate more possible services. Specifically, this time-domain will be used by the DaaS composition algorithm to limit the search and selection scope. The temporal domain is an important component for the DaaS search to enable more relevant services and better performance.
\emph{Stations and Skyways Geo-locating} are processes to call the latitude and longitude coordinates of the delivery stations and Skyways. The outcome of this function will be combined with the \emph{Temporal domain definition} to compute the spatiotemporal attributes. The spatiotemporal attributes will be used to \emph{filter the weather dataset} and construct the network graph.
{The next step is to pass the PDR information to the uncertainty-aware route-planning algorithm as an input.}
\begin{algorithm}[h!]
\caption{DaaS Service Composition}\label{DaaS_Composer}
\begin{algorithmic}[1]
\Function{heuristic}{a,b}
\Comment{The f(h) take two graph nodes: a and b.}
    \State $WS \gets wind speed(a, b) * 3.6 + CM$
    \State $Dist. \gets distance(a, b)$
    \State $WB \gets windBearing(a, b)$
    \State $CB \gets compass\_bearing(a, b)$
    \State $CC \gets cloudCover(a, b)$
    \State $T \gets temperature(a, b) + CM$
    \State $Hu \gets humidity(a, b)$
    \State $V \gets visibility(a, b) + CM$
    \State $DP \gets dew\_point(a, b) - CM$
    \\
    \If{$CC < 0.5$ OR
        $(-20 < T < 45)$ OR
        $ V > 5$ OR
        $Hu < 0.9$ OR $DP < 21$}
      \State $GSA \gets P4\_speed\ OR\ M200\_speed \ 2$
      \State $\Vec{DAS} \gets \Vec{GSA} + \Vec{WS}$
      \State $fd \gets DAS * 60$
      \State {return} $fd$
    \EndIf
\EndFunction

\\
\Function{DaaSComposer}{PDR}
    \State $S   \gets PDR\_source$
    \State $D   \gets PDR\_destination$
    \State $P_t \gets PDR\_pickup\_time$
    \State $P_w \gets PDR\_package\_weight$
    \State $Temporal\_Domain \gets P_t + Time\_Threshold$
    \\
    \State $DaaS\_weather\_P_t \gets weather(P_t)$ 
    \\
    \State $DaaS = nx.Graph()$ 
    \State $DaaS_{nodes} = stations$
    \State $DaaS_{edges} = sky\-ways$
    \State $DaaS\_edges_{attributes} = DaaS\_weather\_P_t$ 
    \\
    \State $Route_{nodes} = astar(DaaS, S, D, Heuristic())$
    \\
    \State $DaaSs = Spatial(DaaSs, Route_{nodes})$
    \State $DaaSs = Temporal(DaaSs, Temporal\_Domain)$
    \State $DaaSs = Weight\-aware(DaaSs, P_w)$
    \State {return} $DaaS_{Composite} \gets DaaSs$
\EndFunction
\end{algorithmic}
\end{algorithm}
\section{Uncertainty-aware Route-Planning Algorithm}\label{route-planning}
{This section discusses the proposed uncertainty-aware route-planning algorithm. First, the network graph is constructed based on the given information by the DaaS scheduling model. Then, the weather inputs are used as the QoS heuristics that guide the route-planning algorithm. Descriptions of the relative airspeed and weather measurements uncertainties are also given in the following subsections.}

\subsection{Finding the Best Route}
The proposed algorithm creates a network graph for each PDR and updates its edge attributes. The algorithm uses the predefined, by the scheduling model, delivery stations and Skyways data to feed the graph nodes and edges, respectively. The algorithm then updates the graph edges attributes with multiple QoS such as distance, temperature, cloud cover, wind speed and bearing, dew-point, and visibility. The algorithm employs the A* (A star) to find the shortest path. A* computes the best route based on the costs of the potential paths and the estimated cost required to extend each path to the destination. It selects the path that minimises the Eq. \ref{Astar} where n is the next node, $g(n)$ is the cost from the start to reach n, and $h(n)$ is a heuristic function which calculates the cost of the cheapest path to reach the destination from n.
\begin{equation}\label{Astar}
  f(n) = g(n) + h(n)
\end{equation}
{The algorithm uses the DaaS QoS attributes to calculate a set of heuristics to find the score of the \emph{h} function.}

\subsection{Weather QoS Properties}
We propose to use the drone capabilities and forecast weather measurements as QoS properties to select a DaaS and Skyway, respectively. The weather conditions of the candidate station must comply with the available drone capabilities to select that station as next fly destination. 
The algorithm then updates the graph edges, i.e. Skyways attributes with  QoS attributes as follows.
\begin{itemize}
\item Distance \emph{Dist.} is the distance between the given two delivery stations.
\item \emph{CC} is the cloud cover percentage. We used a threshold of 0.5\% as a constraint to fly to this node or not. A drone pilot expert defines the utilised threshold.
\item \emph{T} is the temperature that is defined to be between $-20^\circ $ C and $45^\circ $ C according to the drone capabilities listed in Table \ref{drone-types} 
.
\item Humidity \emph{Hu} is the amount of water vapour in the air. \emph{Hu} may cause hardware damage in the long-term. However, \emph{hu} has a less harmful impact on the drone heath in comparison to clouds or temperature. Therefore, we used 0.9 as the threshold of \emph{Hu}. The average of \emph{Hu} in the collected $1,294,185$ weather measurements is 0.6. 
\item  Visibility \emph{V} measures at which distance an object can be detected clearly. Visual positioning and detection are major components in the recently manufactured drones. Visibility statistics are 2, 15, and 10 km for minimum, maximum, and average, respectively. We propose to select DaaS at 5 km to assure that the drone visual algorithm will work efficiently. This threshold complies with \emph{Civil Aviation Legislation Amendment (Part 101) Regulation} which states that the horizontal visibility at least to be 5 kilometres to fly a drone.
\item \emph{WS} is wind speed.
\item \emph{WB} is the wind bearing that will be used to compute the impact of the wind on the drone speed and as a result on the \emph{fd}.
\item  Dew point \emph{DP} measures the temperature to which air must be cooled to become saturated with water vapour. High level of dew point means more moisture in the air. The dew point is related to temperature and humidity. The mean of the dew point measurements of the collected weather data is $7.3^\circ$ C. Severe thunderstorms happen at dew points above $21^\circ $ C. Therefore, we use 21 as a threshold for the selection.
\end{itemize}
We utilise the weather measurements instead of the direct forecast such as \textit{“there will be a thunderstorm”}. The purpose of this process is to produce more accurate calculations. Specifically, we collect the weather measurements for the geo-location coordinates instead of general weather status. We define the thresholds mentioned above according to the aviation safety regulations \footnote{These regulations are different from a country to another. In this research, we follow the regulations that are issued by the Civil Aviation Safety Authority (CASA).} and experts recommendations from online portals \footnote{we use recommendations from websites as the Aircraft Owners and Pilots Association (https://www.aopa.org/) and UAV Forecast (https://www.uavforecast.com/).}
Most of the QoS attributes are used directly to select the best Skyway in the A* algorithm. However, the drone speed requires a prepossessing step to consider the relationship with wind speed and direction.

\subsection{Relative Drone Airspeed}\label{Relative}
\begin{figure}[h!]
\centering
\includegraphics[width=0.3\textwidth]{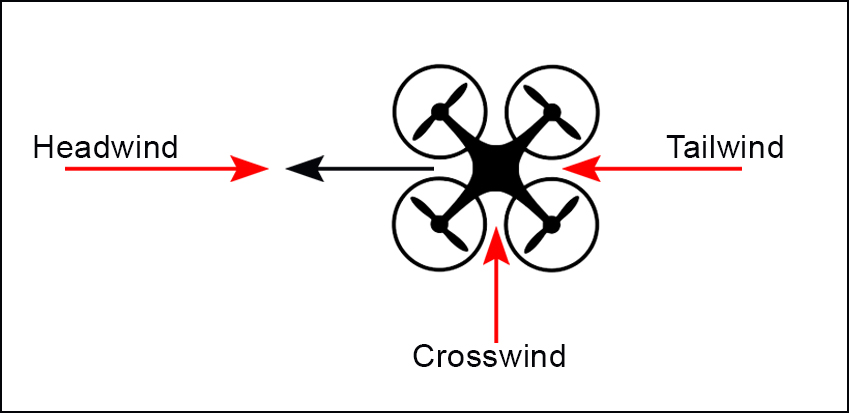}
\caption{Wind types based on the drone and wind bearings.}
\label{Wind-types}
\end{figure}

\begin{figure}[h!]
\centering
\includegraphics[width=0.3\textwidth]{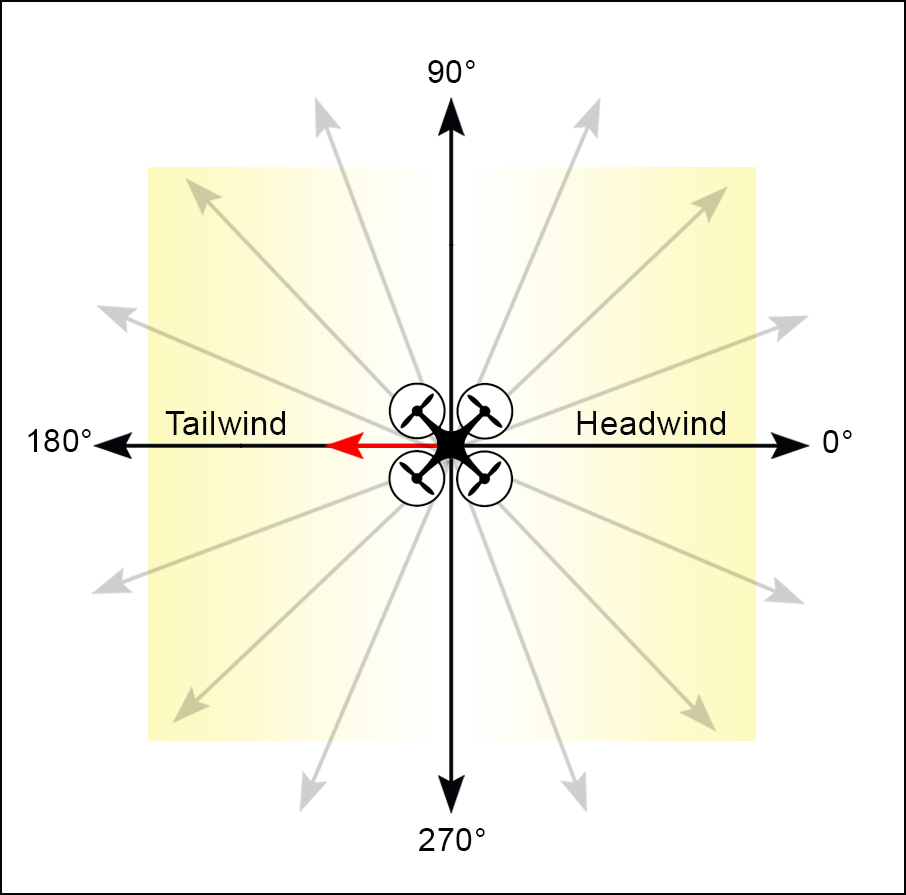}
\caption{Wind type calculation based on the drone and wind bearings.}
\label{Wind-type-calc}
\end{figure}

As a heuristic for A*, we calculate the relative airspeed between the ground speed and the wind speed. At better weather conditions, the drone airspeed is equal to the ground speed. The wind bearing or direction is a dominating factor. There are three different wind types based on the bearing relationship with the flying drone. \emph{Tailwind} happens when the wind and the drone move in the same direction. \emph{Headwind} represents the case when the wind blows against the drone. \emph{Crosswind} happens if the drone bearing is perpendicular to the wind bearing. Tailwind pushes the drone to fly faster and reduce the \emph{fd}. Headwind is preferred in the take-off and landing situations. Crosswind pushes the drone off the planned trajectory. If the crosswind is strong, the drones will have to consume much energy to stay on course and reach the predefined path. Fig. \ref{Wind-types} shows the different wind types according to the relationship between the directions of the drone and the wind. As shown in Fig. \ref{Wind-type-calc}, the angle between the drone direction and wind bearing decides if the wind type is headwind or tailwind. If the angle $90^\circ > \theta > 270^\circ$ refers to a headwind and $90^\circ < \theta < 270^\circ$ refers to a tailwind.
According to the previous rules, the headwind or tailwind and crosswind are calculated as follow:
\begin{equation}\label{"Headwind"}
    ws = windspeed * \cos{(\theta)}
\end{equation}
\begin{equation}\label{"Crosswind"}
    Crosswind = windspeed * \sin{(\theta)}
\end{equation}
The \emph{ws} may be a headwind or a tailwind and $\theta$ is the wind bearing.
We consider calculating the headwind or tailwind to compute the drone airspeed. This process should enable more accurate uncertainty-aware \emph{flight duration} estimation. 
Ground speed average (GSA) is calculated as the average of the utilised drones speeds. The GSA is approximately 77 km/h, where, the GSAs are 72 and 82.8 km/h for the P4 and M200 drones, respectively.
Wind speed of either headwind or tailwind is crawled in the weather dataset and measured with meter per second. We use the following equation to unify the measurement with the GSA to be in km/h:
\begin{equation}\label{"WS"}
    WS = Wind speed * 3.6
\end{equation}
{
Drone air speed (DAS) is computed as the subtraction of the GSA and WS.
\begin{equation}\label{"DAS"}
    \Vec{DAS} = \Vec{GSA} + \Vec{WS}
\end{equation}
This vector addition refers to the adding or subtracting the wind-speed in cases of tail or head wind types.}
Flight duration (fd) is calculated using the Skyways distance and DAS, as follows:
\begin{equation}\label{"ms-khm"}
    fd =  \frac{Distance}{DAS} * 60
\end{equation}
{However, all these calculations depend on the forecast weather measurements. Such measurements suffer from the uncertainty problem. The algorithm selects the best route to contain certain stations based on the currently available forecast weather, at the time of a drone lands at these stations. The actual weather may deviate higher or lower than the previous forecasts.}

\subsection{Uncertainty-aware Model}
To investigate the impact of the uncertainty on the weather measurements, we collected the available historical forecast measurements hourly at the previous $24$ hours of each actual measurement. We collected $364,000$ historical weather data for the utilised $16,000$ weather records, in the small network. For instance, for the actual temperature in station \emph{E} at the timestamp of 11/1/17 1:00, we collected $24$ temperature forecasts at all the previous 24 hours.
We used the historical forecast measurements to compute their deviations from the actual ones. 
We define a Certainty Margin (CM) as two standard deviations of the related measure, e.g., temperature, to be added or subtracted from the forecast measurement. This process aims to offer certain measurements to the DaaS route-planning algorithm to overcome the uncertainty problem. 
Appendix \ref{Appendix_Geo} discusses more explanations on the uncertainty analysis.


\section{DaaS Composition}
{In this section, we explain the proposed DaaS composition methods including the \emph{Spatiotemporal DaaS Composer} and \emph{Predictive DaaS Composer}.}
\subsection{Spatiotemporal DaaS Composer}\label{STcomposer}
\begin{figure}[ht]
\centering
\includegraphics[width=0.42\textwidth]{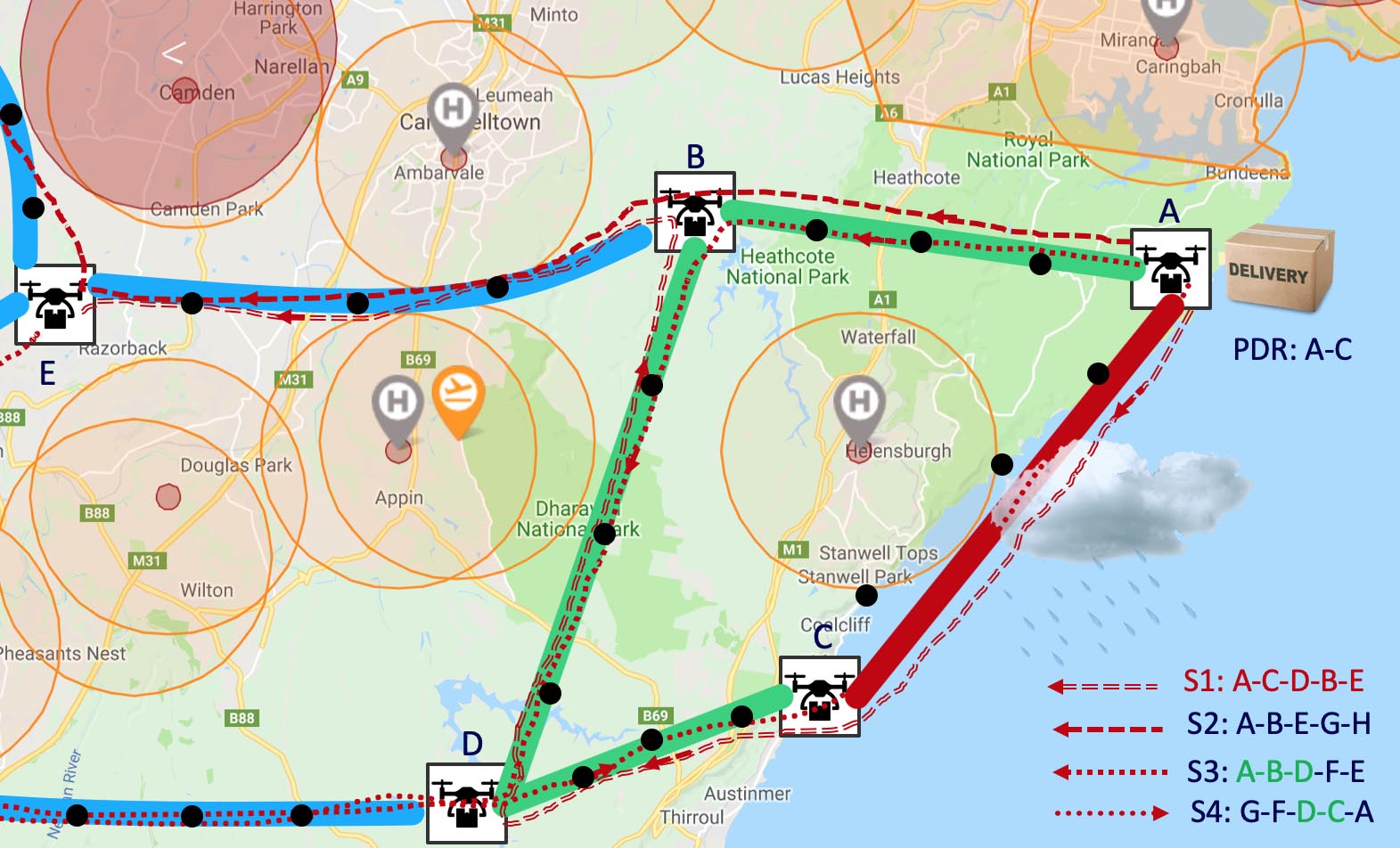}
{\caption{Uncertainty-aware route-planning and service composition. \textit{ Note: we use major stations here for simplicity. This means there are multiple minor stations between each two stations on this map (coloured in black). The detailed map can be found in Fig. \ref{DaaS_network_graph}}}}
\label{DaaS-Composit}
\end{figure}

\begin{figure}[ht]
\centering
\includegraphics[width=0.40\textwidth]{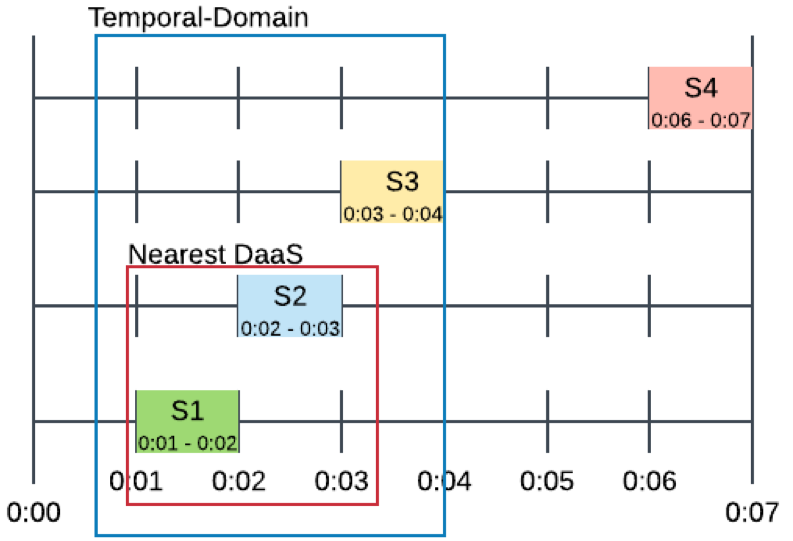}
\caption{An example of the temporal composability between DaaSs.}
\label{DaaS-Temporal}
\end{figure}

The proposed \emph{Spatiotemporal DaaS composer} algorithm is implemented to accomplish the components as shown in Algorithm \ref{DaaS_Composer}. The proposed algorithm receives the PDR information with the selected best route and returns the optimal composite DaaS. 
We define DaaS composite service as a set of DaaSs. DaaSs are considered as composable services if they are connected spatially, found in a particular time domain, and able to handle the PDR package.
As shown in Algorithm \ref{DaaS_Composer}, the proposed algorithm locates each station in the shortest path and compute the three selection functions, including spatiotemporal and weight-aware, to build the optimal composite DaaS delivery service. The algorithm selects the candidate DaaSs spatially based on the computed uncertainty-aware planned route.
Fig. \ref{DaaS-Composit} shows an example of spatial composition based on the outcome of the route-planning algorithm. A PDR requests to deliver a package form station \textit{A} to \textit{C}. Although, there is a direct Skyway from A to C, the uncertainty-aware route-planning algorithm choose to fly through the path \textit{A}, \textit{B}, \textit{D}, and \textit{C}. Then, the algorithm composes DaaS according to the drone payload capacity \emph{pl} and PDR weight.
Then, the algorithm temporally selects the first DaaS available within the predefined temporal domain. Fig. \ref{DaaS-Temporal} presents an example of the temporal composition component. A DaaS S1 is expected to arrive in the time interval from 0:01 to 0:02. The algorithm uses a temporal domain to filter all the available DaaSs. Two DaaSs S2 and S3 are expected to arrive with the temporal domains, and S4 is filtered out.
{The algorithm sorts the spatiotemporal, weight-aware selected DaaS candidates and finally add the time-nearest services to the composite DaaS.}
%
%

{These multiple components of the composition algorithm are expected to produce high-computational cost. Therefore, we propose predictive models to select each service in the DaaS composite service. The next section investigates the utilisation of the outcome of the \emph{spatiotemporal DaaS composer} to extract new features for Machine Learning training.}

\subsection{Predictive DaaS Composer}\label{Pcomposer}
We propose to build an accurate, high-speed DaaS composer using Machine Learning classification models. This proposal is important to overcome the high-cost computations issues of the \emph{spatiotemporal DaaS composer}. This high cost is analysed and discussed later to show the need for the predictive models, refer to Fig. \ref{time_results}. The outcome of the \emph{spatiotemporal DaaS composer} is used as ground-truth to train multiple classification models. The trained model is used to predict the optimal DaaSs instead of the previous spatiotemporal composer. In this scenario, the PDR is processed by the uncertainty-aware route planner, and the predictive composer predicts the optimal DaaS for each station.

{We propose five sets of features to train the classification models. The feature sets are composed of PDR, segment, weather, and spatiotemporal features. The PDR set includes the DaaS request attributes. The segment set includes the current segment properties for each selection process. The weather set includes the previously discussed and collected weather attributes. The spatiotemporal feature set includes new features such as the count of the spatial-available, temporal-available, and weight-aware DaaS candidates.}

\section{Experimental Results}\label{ExpRes}
{This section discusses the experimental results of the proposed uncertainty-aware approach. First, we explain the experiments and results of the route-planning and spatiotemporal composer. Then, we use the outcome of the previous component to train and test the predictive composer.}
\subsection{Route-planning and Spatiotemporal Composition}
We experimented the proposed algorithm on generated 2,000 PDRs. We validated the performance of the proposed algorithm using the uncertainty-aware A* and the greedy algorithm Dijkstra. The optimal DaaSs are composed considering the distance and time based on the drone capabilities and weather conditions. Table \ref{tbl:heuristics} shows two examples of DaaS service composition using A* and Dijkstra. The first row presents the DaaS composite service, which is found as for the PDR number 3. Dijkstra utilised its greedy methodology to select the shortest route with seven services, 72 km, and 54 minutes. On the other hand, the uncertainty-aware algorithm using A* results in different distances and flight durations, as can be seen in Fig. \ref{DaaS_resutls_error_rates}. The repetition in the DaaS services IDs in Table \ref{tbl:heuristics} means that this DaaS service is consumed for two consecutive Skyway segments.  
{
\begin{table*}[]
\caption{Experimental results show the impact of utilised heuristics showing significant changes due to the uncertainty awareness.}
\label{tbl:heuristics}
\begin{tabular}{|c|c|c|c|p{3.33cm}|c|c|c|p{3.33cm}|}
\hline
\multirow{2}{*}{\begin{tabular}[c]{@{}c@{}}PDR\\ ID\end{tabular}} & \multicolumn{4}{c|}{Dijkstra} & \multicolumn{4}{c|}{A*} \\ \cline{2-9} 
 & \begin{tabular}[c]{@{}c@{}}DaaS\\ Count\end{tabular} & \begin{tabular}[c]{@{}c@{}}Distance\\ km\end{tabular} & \begin{tabular}[c]{@{}c@{}}Duration\\ Minutes\end{tabular} & \begin{tabular}[c]{@{}c@{}}DaaS Services \\ IDs\end{tabular} & \begin{tabular}[c]{@{}c@{}}DaaS\\ Count\end{tabular} & \begin{tabular}[c]{@{}c@{}}Distance\\ km\end{tabular} & Duration Minutes & \begin{tabular}[c]{@{}c@{}}DaaS Services \\ IDs\end{tabular} \\ \hline
3 & 7 & 72 & 54 & 6, 9, 16, 58, 18, 26 & 8 & 84 & 62 & 6, 9, 18, 18, 58, 71, 16 \\ \hline
45 & 8 & 64 & 51 & 575, 663, 663, 859, 686, 872, 809 & 9 & 97 & 73 & 575, 663, 663, 859, 721, 758, 542, 646  \\ \hline
\end{tabular}
\end{table*}
}
{
\begin{table}[]
\caption{The impact of the uncertainty-aware A* against the greedy algorithm on the composed DaaS services.}
\label{resutls-counts}
\begin{tabular}{|l|p{2.4cm}|l|l|}
\hline
 & Equal & A* more & Dijkstra more \\ \hline
DaaS counts & 1589 & 411 & 0 \\ \hline
Distances & 1413 & 360 & 227 \\ \hline
Flying durations & 1384 & 322 & 294 \\ \hline
\end{tabular}
\end{table}
}

We evaluate the significance of the proposed uncertainty-aware algorithm based on three factors. First, we calculated the count of the DaaS selections in the composite DaaS. For instance, Table \ref{tbl:heuristics}, the DaaS service IDs represent the selected services in the final composite DaaS. Here, the results show the impact of the uncertainty-awareness on the DaaS selections. Second, we computed the total distances that are flown by the DaaSs. Third, we estimated the total flying durations in minutes. Fig. \ref{DaaS_resutls_error_rates} highlights how many identical or different DaaSs between the three factors using the two algorithms, i.e., A* and Dijkstra. Identical means that the A* and Dijkstra have the same selected component DaaSs. 
Fig. \ref{DaaS_resutls_error_rates} shows an analysis of the uncertainty impacts on the selections of both A* and Dijkstra. The sub-graph named \emph{count of DaaSs} in Fig. \ref{DaaS_resutls_error_rates} shows the percentages of how identical or different the heuristic A* and Dijkstra are in terms of services count. The results show that the two algorithms have only $10\%$ same count of selected component drone services in the composite DaaS. In the $90\%$ difference means that in around $1800$ PDRs, the proposed uncertainty-aware using A* has different selections. However, this big difference in the selected services does not significantly affect the total distances and flying durations. Fig. \ref{DaaS_resutls_error_rates} shows that the differences are $29.4$ and $30.8$ percentages for the distances and durations, respectively. These statistics prove the efficiency of the proposed uncertainty-aware A* algorithm. It selects different DaaSs while having the same distances and durations in $70\%$ of the total PDRS.

\begin{figure}[h!]
\centering
\includegraphics[width=.45\textwidth]{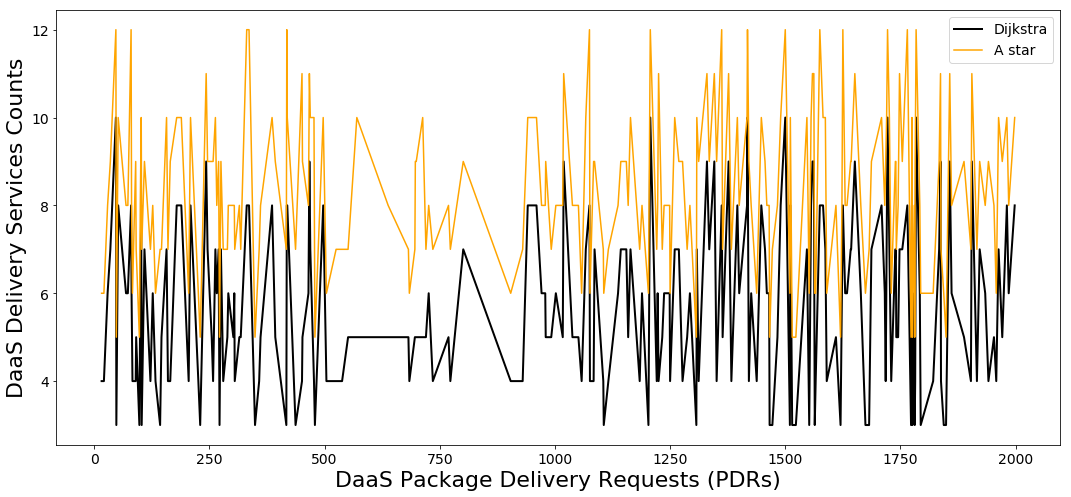}
\caption{The count of selected services in each composite DaaS with at least 2 services difference between A* and Dijkstra.}
\label{DaaS_resutls_lengths}
\end{figure}

Fig. \ref{DaaS_resutls_lengths} highlights the count of selected services in each composite DaaS. The statistics in the figure are calculated for composite DaaSs with at least two services difference between the two algorithms. Using the A* with the proposed heuristics tends to compose the DaaS service with more services than using the greedy algorithm. This is an expected result due to the fact that the less number of the DaaS services may not lead to the optimal DaaS composite service. In such cases, the greedy algorithm may lead to service failures due to ignoring the weather conditions and drone capabilities.   

\begin{figure}[h!]
\centering
\includegraphics[width=.45\textwidth]{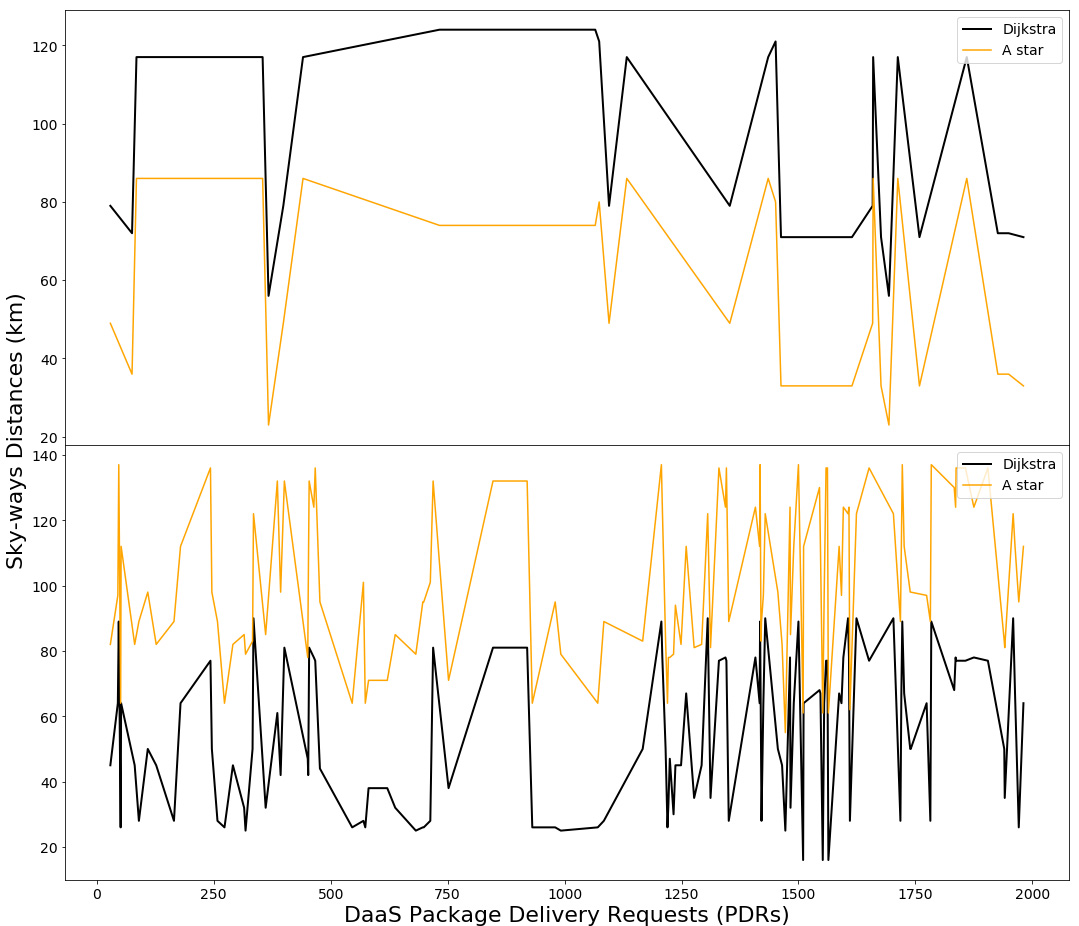}
\caption{Selections in composite DaaS with at least 30km.}
\label{DaaS_resutls_distances_A}
\end{figure}

We analyse the total distances of the flights' Skyways in the final composite DaaS services.
Fig. \ref{DaaS_resutls_distances_A} visualises the differences above 30 km between the composite DaaS services distances for the uncertainty-aware A* and Dijkstra, respectively. The $30$ km threshold is chosen to minimise the data points on the graph for better visualisation.

{Fig. \ref{DaaS_resutls_error_rates} shows about $30\%$ difference between the number of selections of two algorithms in the $2,000$ PDRs. As Dijkstra is a  greedy algorithm which selects the fewest possible number of Skyways services, their distances are shorter than the A* algorithm. As listed in Table \ref{resutls-counts}, the uncertainty-aware A* selects services with longer distances in $360$ composite DaaS. The Dijkstra-based compositions are longer in distances at $227$ DaaSs. This means that Dijkstra compositions are less in only $133$ services or about $0.06\%$. These results prove that the efficiency of the proposed composition algorithm. Such efficiency is not only in terms of the weather conditions and drone capabilities but also searching for the possible shortest path.}

\begin{figure}[h!]
\centering
\includegraphics[width=.45\textwidth]{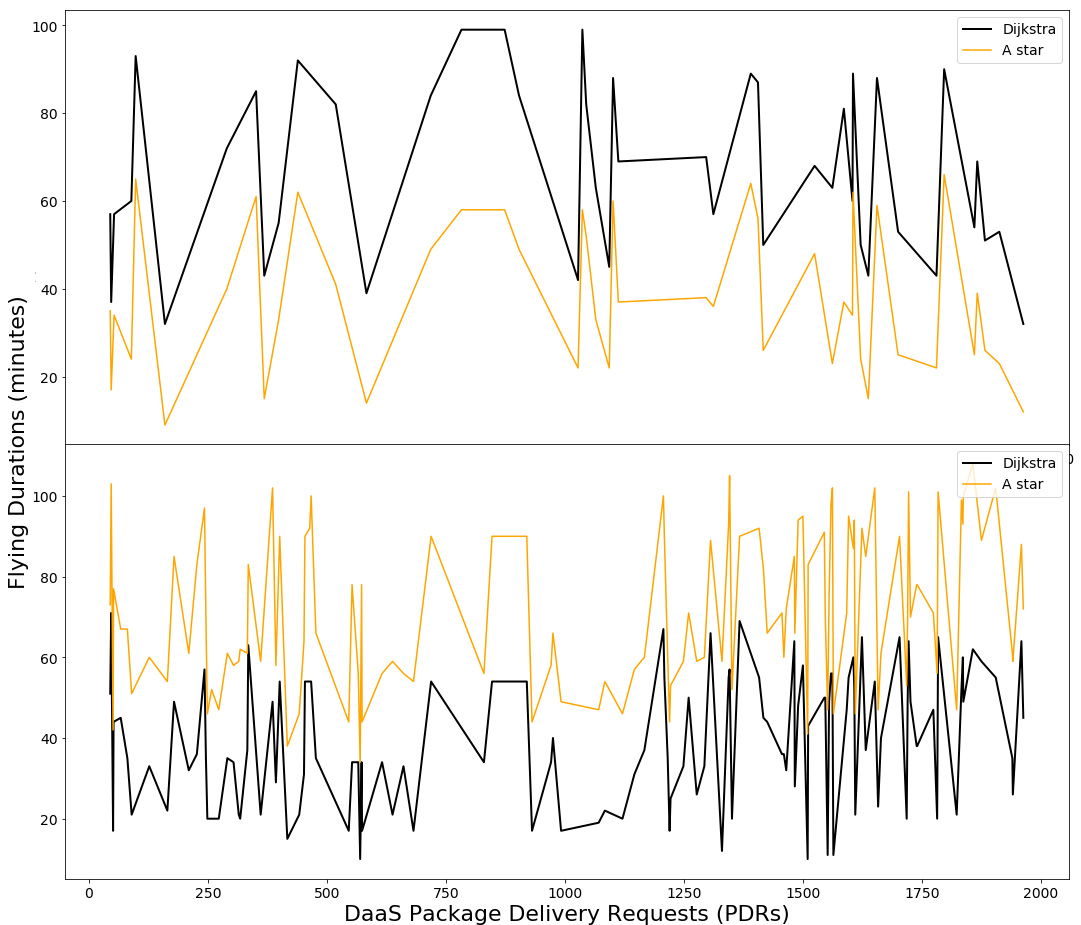}
\caption{Selections in composite DaaS with at least 20 minutes.}
\label{DaaS_resutls_flying_durations_A}
\end{figure}

{Fig. \ref{DaaS_resutls_flying_durations_A} compares experimental results in terms of total flying durations (at least 20 minutes) for the uncertainty-aware A* and Dijkstra algorithms, respectively. Consistently with the increase of the total distances, the total flying durations are increased. Even though the proposed uncertainty-aware A* produces longer durations in only $28$ ($0.01\%$) out of the $2,000$ DaaSs. This result emphasises more on the efficiency of our proposed algorithm. However, it selects different DaaS in 90\% of the compositions; it selects the best few durations. The reason behind this few duration is that the algorithm is aware of the drone speed and select the shortest path based on the flight durations. These flight durations are calculated based on the relative drone airspeed, as discussed in Section \ref{Relative}.}

Here, we also notice that the generated DaaS services were successfully processed by the proposed DaaS framework. This dataset is generated based on the combination of the real weather data, real-coordinates for the delivery stations and Skyways, and the real capabilities of drone models. The spatial distribution of the delivery stations are constrained by the CASA regulations as in Fig. \ref{DaaS_network_graph}. As a prototype design, we have also selected to make it far from the populous areas to avoid the privacy issues.

\begin{table*}[]
\caption{Evaluation of the proposed DaaS classification models using different feature sets.}\label{tbl:ML_results}
\begin{tabular}{|p{1.2cm}|l|l|l|l|l|l|l|l|l|p{1.2cm}|p{1.5cm}|p{1cm}|}
\hline
\multirow{2}{1cm}{Feature Set} &
  \multirow{2}{.5cm}{Model} &
  \multicolumn{8}{c|}{\begin{tabular}{c}\\Small Network (38 nodes)\end{tabular}} &
  Large Network (1254 nodes) &
  \multirow{2}{*}{Features} &
  \multirow{2}{0.6cm}{Feature count} \\ \cline{3-11}
 &
   &
  CM=1 &
  CM=2 &
  CM=3 &
  CM=4 &
  CM=5 &
  CM=10 &
  CM=15 &
  CM=20 &
  CM=2 &
   &
   \\ \hline
\multirow{4}{*}{X1} &
  SGD &
  37.56\% &
  38.50\% &
  37.17\% &
  35.10\% &
  33.62\% &
  33.58\% &
  33.56\% &
  37.28\% &
  41.26\% &
  \multirow{4}{*}{PDR} &
  \multirow{4}{*}{7} \\ \cline{2-11}
 &
  SVM &
  37.62\% &
  34.52\% &
  30.15\% &
  31.62\% &
  32.85\% &
  25.42\% &
  28.70\% &
  29.01\% &
  35.68\% &
   &
   \\ \cline{2-11}
 &
  LR &
  49.34\% &
  49.44\% &
  49.29\% &
  46.21\% &
  44.64\% &
  45.64\% &
  45.38\% &
  44.96\% &
  42.31\% &
   &
   \\ \cline{2-11}
 &
  GNB &
  51.23\% &
  50.68\% &
  50.08\% &
  43.33\% &
  41.57\% &
  40.44\% &
  38.61\% &
  39.81\% &
  44.9\% &
   &
   \\ \hline
\multirow{4}{*}{X2} &
  SGD &
  88.83\% &
  88.15\% &
  85.89\% &
  78.23\% &
  77.74\% &
  73.50\% &
  72.30\% &
  70.92\% &
  64.52\% &
  \multirow{4}{1.5cm}{PDR + segment} &
  \multirow{4}{*}{15} \\ \cline{2-11}
 &
  SVM &
  80.86\% &
  81.23\% &
  78.52\% &
  75.17\% &
  71.76\% &
  67.06\% &
  67.56\% &
  66.81\% &
  50.28\% &
   &
   \\ \cline{2-11}
 &
  LR &
  93.23\% &
  93.34\% &
  93.12\% &
  85.29\% &
  83.63\% &
  82.30\% &
  82.80\% &
  81.76\% &
  61.33\% &
   &
   \\ \cline{2-11}
 &
  GNB &
  95.82\% &
  95.83\% &
  95.21\% &
  88.30\% &
  84.93\% &
  83.39\% &
  80.35\% &
  79.82\% &
  84.54\% &
   &
   \\ \hline
\multirow{4}{*}{X3} &
  SGD &
  97.82\% &
  98.17\% &
  97.98\% &
  91.64\% &
  91.55\% &
  89.17\% &
  93.82\% &
  90.58\% &
  98.94\% &
  \multirow{4}{1.5cm}{PDR + segment + weather} &
  \multirow{4}{*}{17} \\ \cline{2-11}
 &
  SVM &
  97.12\% &
  95.77\% &
  97.22\% &
  90.47\% &
  90.44\% &
  87.02\% &
  90.68\% &
  90.65\% &
  97.66\% &
   &
   \\ \cline{2-11}
 &
  LR &
  98.81\% &
  98.92\% &
  99.09\% &
  94.54\% &
  94.24\% &
  91.68\% &
  93.94\% &
  94.18\% &
  99.6\% &
   &
   \\ \cline{2-11}
 &
  GNB &
  96.61\% &
  96.53\% &
  96.49\% &
  91.47\% &
  88.89\% &
  88.01\% &
  88.26\% &
  88.28\% &
  97.63\% &
   &
   \\ \hline
\multirow{4}{*}{X4} &
  SGD &
  94.23\% &
  94.49\% &
  94.15\% &
  83.14\% &
  86.40\% &
  80.76\% &
  82.09\% &
  82.56\% &
  86.19\% &
  \multirow{4}{1.5cm}{PDR + segment + spatiotemporal} &
  \multirow{4}{*}{11} \\ \cline{2-11}
 &
  SVM &
  89.45\% &
  84.50\% &
  89.49\% &
  83.56\% &
  80.60\% &
  79.73\% &
  78.11\% &
  79.67\% &
  83.62\% &
   &
   \\ \cline{2-11}
 &
  LR &
  96.05\% &
  96.15\% &
  95.77\% &
  89.85\% &
  88.38\% &
  88.54\% &
  89.49\% &
  89.24\% &
  85.32\% &
   &
   \\ \cline{2-11}
 &
  GNB &
  98.05\% &
  98.04\% &
  97.30\% &
  92.41\% &
  90.48\% &
  88.42\% &
  85.38\% &
  84.57\% &
  95.19\% &
   &
   \\ \hline
\multirow{4}{*}{X5} &
  SGD &
  98.39\% &
  98.72\% &
  98.46\% &
  91.98\% &
  92.11\% &
  94.10\% &
  94.19\% &
  93.42\% &
  99.71\% &
  \multirow{4}{1.9cm}{PDR + segment + spatiotemporal + weather} &
  \multirow{4}{*}{19} \\ \cline{2-11}
 &
  SVM &
  97.47\% &
  97.36\% &
  97.49\% &
  92.59\% &
  92.70\% &
  93.17\% &
  94.60\% &
  94.63\% &
  99.79\% &
   &
   \\ \cline{2-11}
 &
  {LR} &
  {99.74\%} &
  {99.84\%} &
  {99.68\%} &
  {95.72\%} &
  {95.75\%} &
  {95.83\%} &
  {96.02\%} &
  {95.95\%} &
  {99.89\%} &
   &
   \\ \cline{2-11}
 &
  GNB &
  97.15\% &
  97.15\% &
  96.90\% &
  93.05\% &
  90.68\% &
  89.55\% &
  89.94\% &
  90.24\% &
  98.42\% &
   &
   \\ \hline
\end{tabular}
\end{table*}

\begin{figure}[h]
\centering
\includegraphics[width=0.43\textwidth]{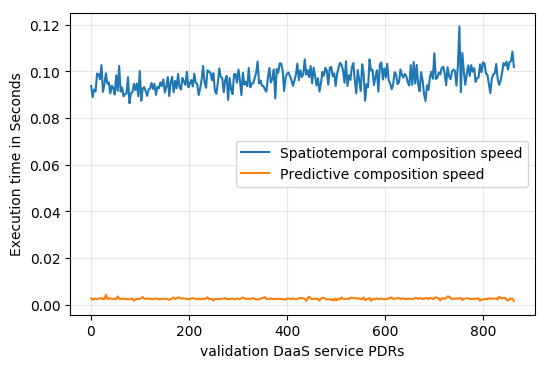}
\caption{A comparison of the computation cost between the spatiotemporal and the predictive DaaS composition methods.}
\label{time_results}
\end{figure}
\begin{figure*}[!ht]
\centering
\includegraphics[width=0.99\textwidth]{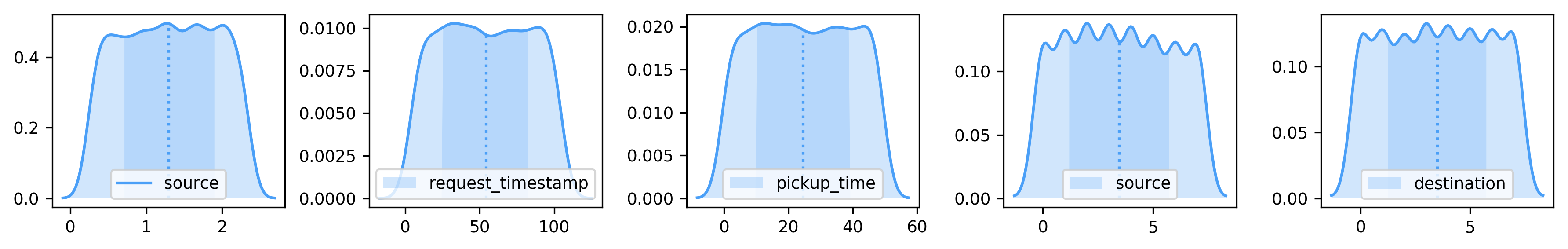}
\caption{DaaS delivery requests show uni-from distribution over all the data attributes.}
\label{PDR_dist2}
\end{figure*}
\subsection{Prediction Experimental Results}
We generate a new dataset of $5,000$ PDRs to deliver packages among some sources and destinations around the delivery station \emph{E}. We choose \emph{E} because it is located in the centre of the network and has a large number of DaaS services, please refer to Fig. \ref{DaaS_distribtions}.
{Fig. \ref{PDR_dist2} highlights the DaaS delivery requests distributions. The delivery requests generated data show uni-from distribution over all the data attributes. The DaaS requests are symmetrically distributed over delivery weight, request time, geo-locations of source and destination, and pick-up time. More analyses on the generated data distributions are found in Appendix \ref{Appendix_data}.}
We run the above-discussed components, as in Fig. \ref{DaaS-approach}, and collected the final selections by the \emph{spatiotemporal DaaS composer}. This process results in a large number of DaaS selections for the selected network stations.

{The generated $5,000$ PDRs dataset results in $1,256,136$ and $25,676$ selections that are made by the \emph{spatiotemporal DaaS composer} using the large and small networks, respectively.}
First, we utilised the small network to experiment with the different CM values. We then run the learning experiment on the large network with best CM value.
The selections are used as observations for cross-validation.  
We used the Python Scikit-learn package to implement the predictive DaaS composition models \cite{pedregosa2011scikit}. We trained four classifiers that represent different Machine Learning methods. The trained classifiers are Stochastic Gradient Descent (SGD), Support Vector Machine (SVM), Logistic Regression (LR), and Gaussian Naive Bayes (GNB).
Table \ref{tbl:ML_results} shows the results of the cross-validation using the feature sets from the large and small network, including $X1$, $X2$, $X3$, $X4$, and $X5$. $X1$ consists of the PDR features only. X2 combines the PDR and segments features. X3 adds the weather features to $X2$. $X4$ combines $X2$ with spatiotemporal features. Finally, X5 contains all four feature types. 
The weather properties improved the classification results as can be seen in the cases of $X3$ and $X5$. However, the new computed spatiotemporal features outperform the weather features. This is because the weather data are collected hourly while the spatiotemporal features are calculated for each timestamp at each delivery station. The small network based results show that using the {LR} classifier with the feature set $X5$ tend to the best performance with $99.84\%$ for $10-folds$ cross-validation using the $CM2$. However, the large network data lead to a more accurate selection of $99.89\%$.

In the last set of experiments, we evaluated the computational cost of the two proposed methods, i.e., the \emph{spatiotemporal DaaS composer} and \emph{predictive DaaS composer}. Fig. \ref{time_results} shows a comparison between their computation costs. The average execution times are ~0.01 and ~0.002 second for the \emph{spatiotemporal DaaS composer} and \emph{predictive DaaS composer}, respectively. The superiority of the \emph{predictive DaaS composer} proves our hypothesis in using the machine learning to make the DaaS composition faster.

\begin{figure}[h]
\centering
\includegraphics[width=0.43\textwidth]{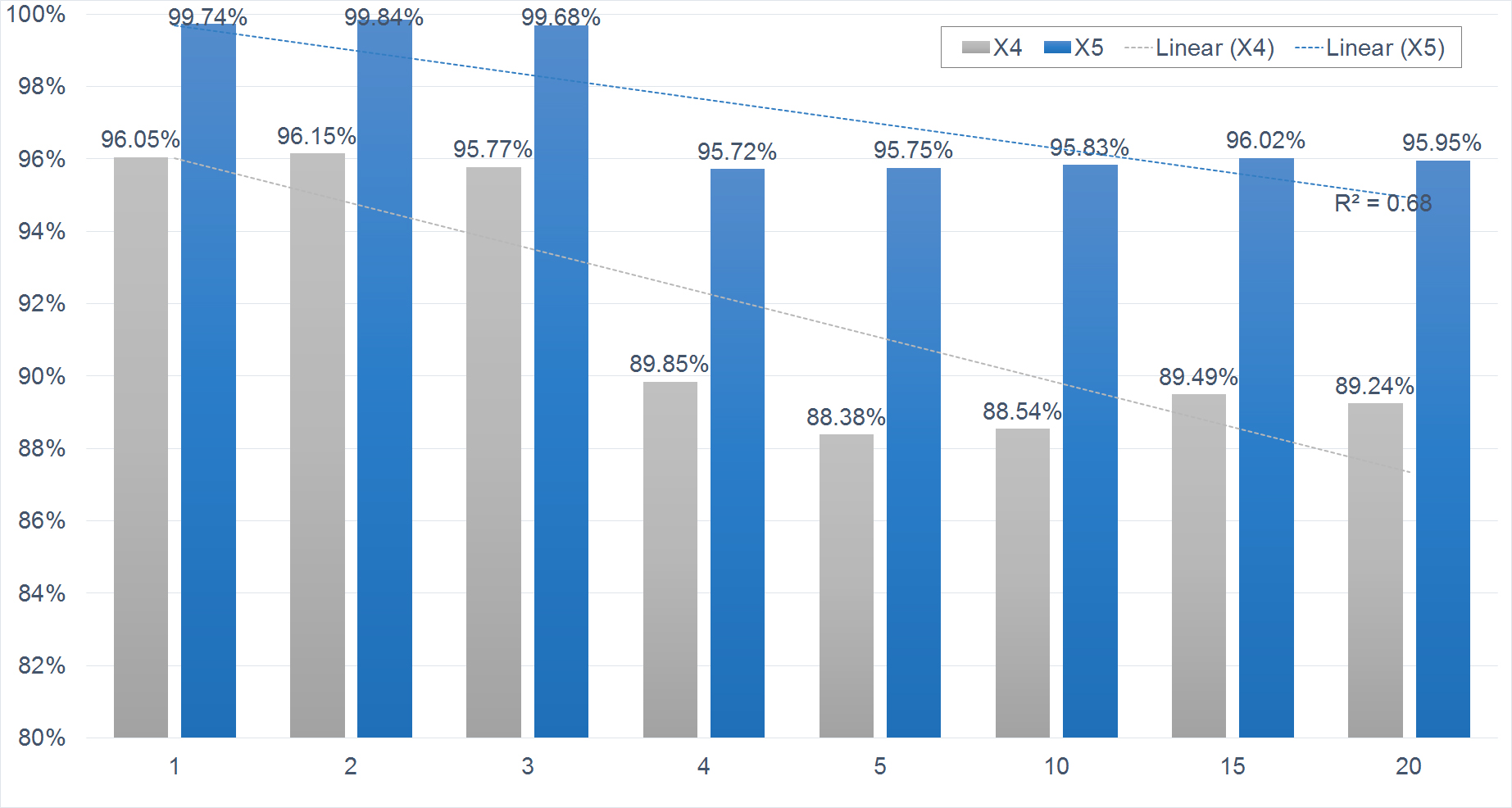}
\caption{The impact of using different CM values on the Logistic Regression classifier performance using the small network dataset.}
\label{CM_learning}
\end{figure}

\begin{figure}[ht]
\centering
\includegraphics[width=.43\textwidth]{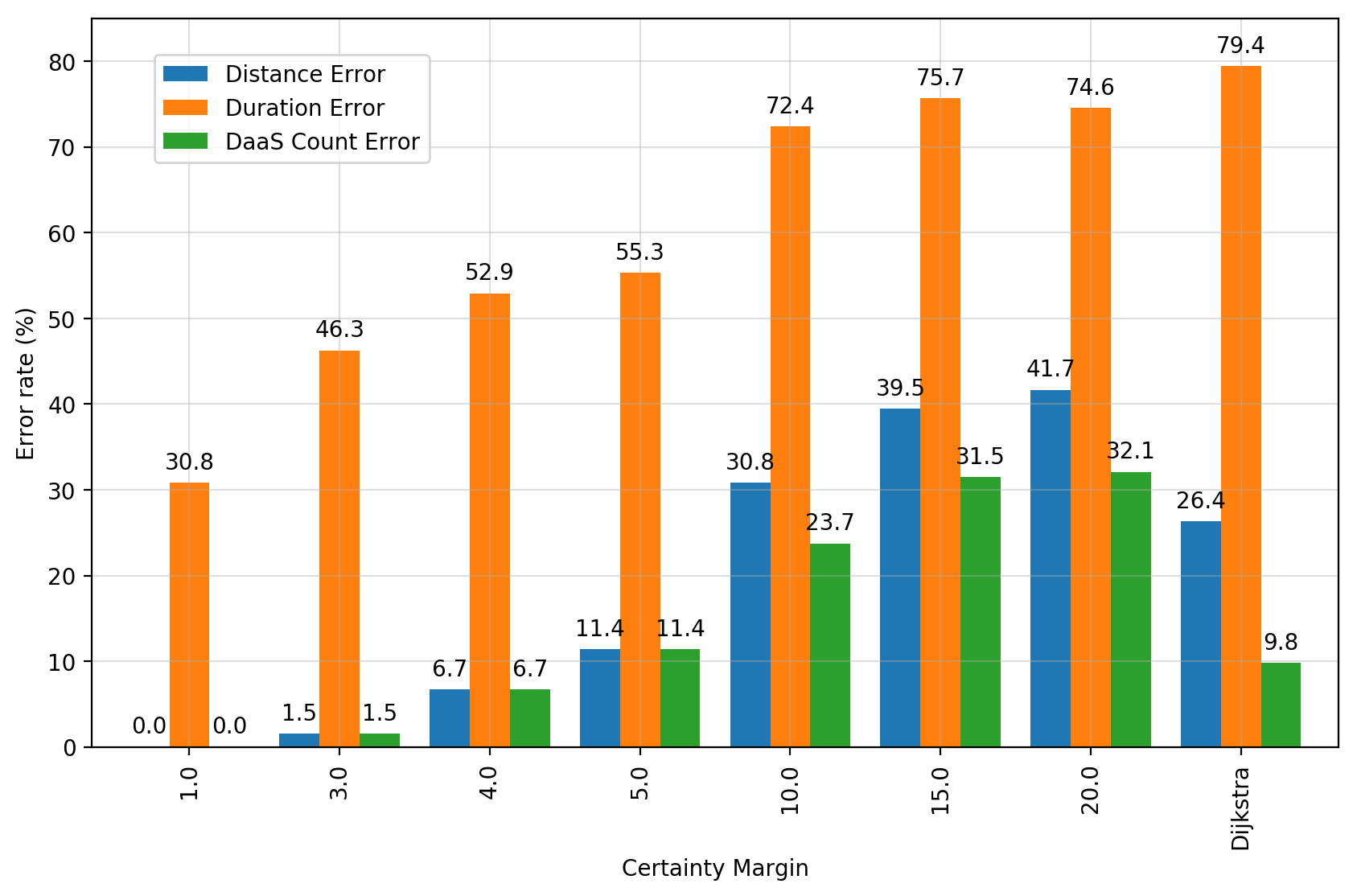}\vspace{-4mm}
\caption{Error percentages of using different CMs in comparison to the proposed $CM2$.}\vspace{-4mm}
\label{DaaS_resutls_error_rates}
\end{figure}

{
\subsection{Error Analysis}\label{Error_Analysis}
In this section, we investigate the impact of using different CM values on machine learning performance.
We have selected the CM value to be a vector addition of two standard deviations ($CM2$) of the related measure, e.g., temperature. This selection is made based on the outcome of the error analysis on using different CM values such as one, two, three and more, as in Fig. \ref{CM_learning} and \ref{DaaS_resutls_error_rates}.
Fig. \ref{CM_learning} shows the experimental results of using different CM values to compute the DaaS features for training the Logistic Regression classifier. Using the model with two standard deviations outperforms the other CMs. Using one standard deviation ($CM1$) comes in second place with slightly decreased accuracy. Fig. \ref{CM_learning} shows the accuracy of using the $X4$ and $X5$ feature sets. It also shows a best-fit straight linear trend-line that describes the accuracy decreasing after the two standard deviation CM.
Based on this outcome, we define the $CM2$ to be the best scheme in order to compare the different CM values. As using different CMs proves to have different learning performance, we investigate the CM impact on having wrong DaaS selections that have led to that accuracy decreasing. Fig. \ref{DaaS_resutls_error_rates} compares the different CMs to the $CM2$ in terms of their composite DaaS distance, duration, and length error. The $CM1$ has zero distance and length error while having 30.8\% error in the duration calculation in comparison to the behaviour of the $CM2$. The error percentages of the three measures are steadily increasing when we increase the CM value. The $CM20$, for example, has $41.7\%$, $74.6\%$, and $32.1\%$ error selection for the distance, flying duration, and the DaaS length, in comparison to the $CM2$.
We have also compared the utilisation of Dijkstra/A-star without the proposed heuristic algorithm. Fig. \ref{DaaS_resutls_error_rates} shows that Dijkstra has high error rates in comparison to the proposed heuristic A-star with the $CM2$.
}


\section{Conclusion }\label{conclusion}
{We introduced a novel uncertainty-awareness framework to deploy DaaS. We define the DaaS service functions and QoS properties to find the best DaaSs. We augment a real Skyways network with synthetic data of the drone itineraries and delivery requests and real-time weather data. We propose a new \emph{uncertainty-aware DaaS route-planning} based on A* algorithm considering the spatiotemporal attributes, drone capabilities, and weather conditions. The planned route is utilised \emph{spatiotemporal DaaS composer} to compose the optimal DaaS form the scheduled DaaS itineraries. Finally, we introduce a new \emph{predictive DaaS composer} to produce fast, accurate DaaS compositions. In future works, there are multiple interesting directions to investigate, such as the DaaS schedule conflicts and the uncertainty of drone battery consumption.}\vspace{-4mm}

\ifCLASSOPTIONcompsoc
  \section*{Acknowledgments}
\else
  \section*{Acknowledgment}
\fi
Ali Hamdi is supported by RMIT Research Stipend Scholarship and the Australian Research Council's Linkage Projects funding scheme (project LP150100246). This work has been supported in part by the ARC Discovery and LIEF programs (DP160103595 and LE180100158).\vspace{-4mm}

\bibliographystyle{IEEEtran}
\bibliography{DaaSref}

\begin{IEEEbiography}[{\resizebox{1in}{!}{\includegraphics{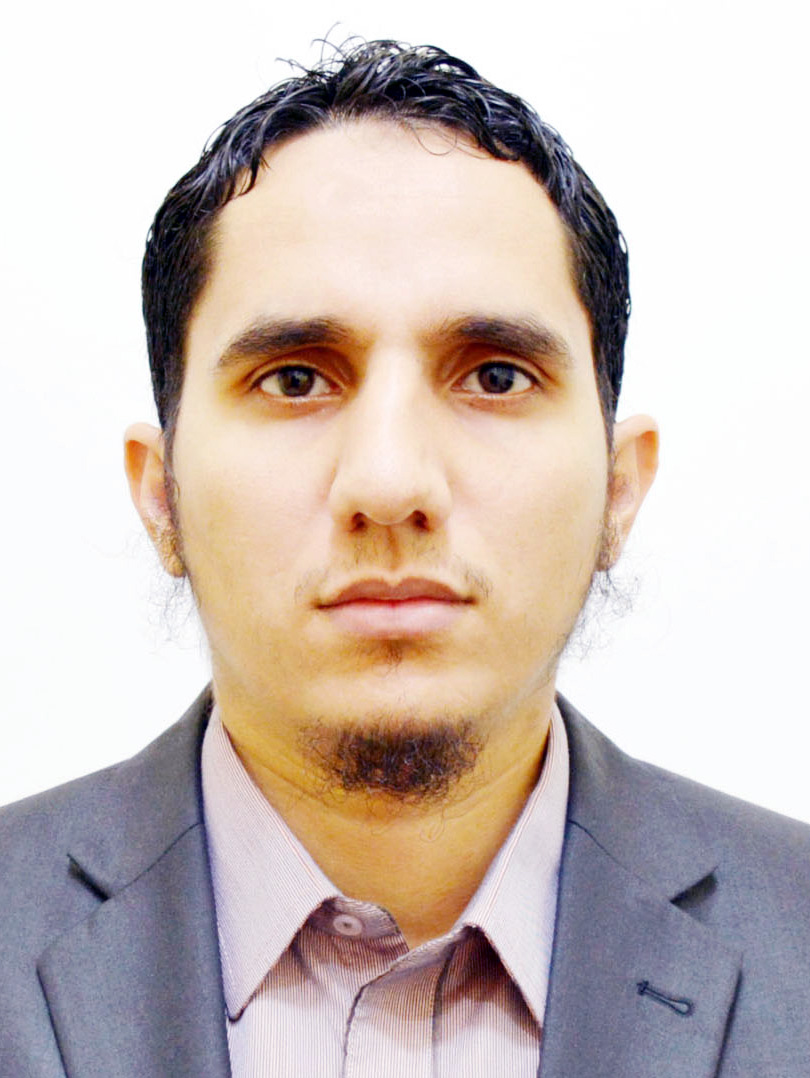}}}]{Ali Hamdi}
is a Ph.D. student in Computer Science at the School of Computing Technologies, RMIT University. He received his master in Computing from University of Technology, Malaysia (UTM) in 2017. He worked as a Data Scientist and Research Assistant. Ali has also taught Computer Programming, Data Science, and Web Development courses. He has participated in many software developments and scientific research projects including, machine learning, natural language processing, image processing, and spatiotemporal data analysis. Ali has worked on drone-based visual object tracking and crowed counting and drone-as-a-service.
\end{IEEEbiography}

\begin{IEEEbiography}[{\resizebox{1in}{!}{\includegraphics{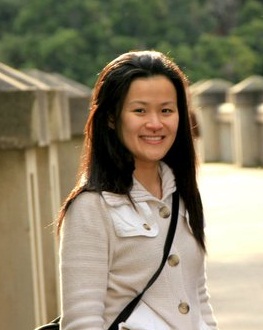}}}]{Flora D. Salim} is a Professor at the School of Computing Technologies, RMIT University. She received a BComp(Hons) with First Class Honours in 2003 and Ph.D. in 2009, both from Monash University. She is the co-Deputy Director of the RMIT Centre for Information Discovery and Data Analytics. She leads the Context Recognition and Urban Intelligence (CRUISE) group. Her research interests include Time-Series, Spatio-Temporal and Trajectory Data Mining, Context and Behaviour Modelling, Unsupervised and Self-supervised Learning. She has received multiple awards, including Humboldt-Bayer Fellowship, Alexander von Humboldt Fellowship (2019), Victoria Fellow 2018, RMIT Award for Research Impact - Technology 2018; RMIT Vice-Chancellor's Award for Research Excellence – Early Career Researcher 2016; Victorian iAwards 2014, Australian Research Council (ARC) Postdoctoral Research Industry Fellow 2012-2015; IBM Smarter Planet Innovation Award 2010; Google Anita Borg Scholar 2008. She is an Associate Editor of PACM IMWUT and an Area Editor of Pervasive and Mobile Computing.
\end{IEEEbiography}

\begin{IEEEbiography}[{\resizebox{1in}{!}{\includegraphics{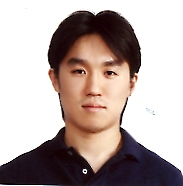}}}]{Du Yong Kim}
received the B.E. degree in electrical and electronics engineering from Ajou University, Korea, in 2005. He received the M.S. and Ph.D. degrees from the Gwangju Institute of Science and Technology, Korea, in 2006 and 2011, respectively. As a Postdoctoral Researcher, he worked on statistical signal processing and image processing at the Gwangju Institute of Science and Technology (2011–-2012), the University of Western Australia (2012–-2014), and Curtin University (2014--2018). He is currently working as a Vice-Chancellor's Research Fellow at School of Engineering, RMIT University. His main research interests include Bayesian filtering theory and its applications to machine learning, computer vision, sensor networks, and automatic control.
\end{IEEEbiography}

\begin{IEEEbiography}[{\resizebox{1in}{!}{\includegraphics{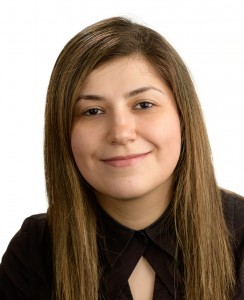}}}]{Azadeh Ghari Neiat}
is a lecturer in the School of Information Technology at the Deakin University, VIC, Australia. Before joining Deakin University, she was a postdoctoral research fellow at the University of Sydney. She was awarded a PhD in computer science at RMIT University, Australia in 2017. Her current research interests include Internet of Things (IoT), Spatiotemporal Data Analysis, Mobile Crowdsourcing/Crowdsensing, Big Data Mining, and Machine Learning with applications in the smart city, smart home, and recommender systems.
\end{IEEEbiography}

\begin{IEEEbiography}[{\resizebox{1in}{!}{\includegraphics{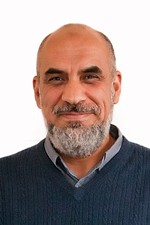}}}]{Athman Bouguettaya} is a Professor of Computer Science at The University of Sydney, NSW, Australia. He received his PhD in Computer Science from the University of Colorado at Boulder (USA) in 1992. He was previously Professor and Head of School of Computer Science and Information Technology at RMIT University, Melbourne, Australia and Science Leader in Service Computing at CSIRO ICT Centre, Canberra. Australia. Before that, he was a tenured faculty member and Program director in the Computer Science department at Virginia Polytechnic Institute and State University (commonly known as Virginia Tech) (USA).  He is a founding member and past President of the Service Science Society, a non-profit organization that aims at forming a community of service scientists for the advancement of service science. He is or has been on the editorial boards of several journals including, the IEEE Transactions on Services Computing, ACM Transactions on Internet Technology, the International Journal on Next Generation Computing, VLDB Journal, and Distributed and Parallel Databases Journal. He served as a guest editor of a number of special issues including the special issue of the ACM Transactions on Internet Technology on Semantic Web services, a special issue the IEEE Transactions on Services Computing on Service Query Models, and a special issue of IEEE Internet Computing on Database Technology on the Web. He served as a Program Chair of the 2012 International Conference on Web and Information System Engineering, the 2009 and 2010 Australasian Database Conference, 2008 International Conference on Service Oriented Computing (ICSOC) and the IEEE RIDE Workshop on Web Services for E-Commerce and E-Government (RIDE-WS-ECEG’04). He has published more than 200 books, book chapters, and articles in journals and conferences in the area of databases and service computing (e.g., the IEEE Transactions on Knowledge and Data Engineering, the ACM Transactions on the Web, WWW Journal, VLDB Journal, SIGMOD, ICDE, VLDB, and EDBT). He was the recipient of several federally competitive grants in Australia (e.g., ARC) and the US (e.g., NSF, NIH). He is a Fellow of the IEEE and a Distinguished Scientist of the ACM.
\end{IEEEbiography}

\newpage
\onecolumn
\setcounter{page}{1}
\setcounter{section}{0}
\setcounter{figure}{0}
\setcounter{table}{0}

\section*{Supplemental Material}
This document contains supplemental material for the TSC paper "Drone-as-a-Service Composition Under Uncertainty".

\begin{appendices}



\section{Geo Visualisation for the Weather Uncertainty}\label{Appendix_Geo}

\begin{figure*}[!ht]
\centering
\includegraphics[width=0.99\textwidth]{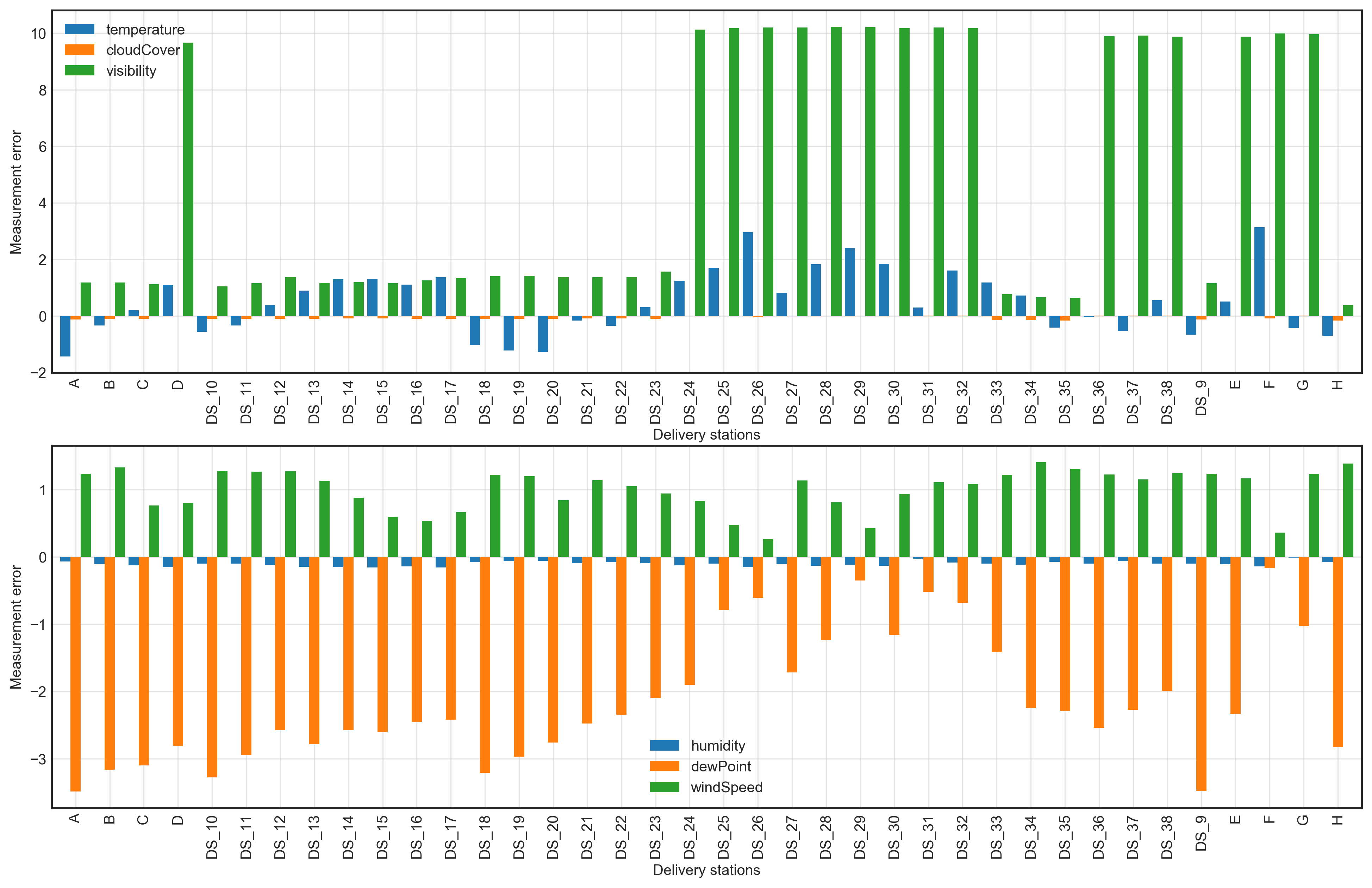}
\caption{The uncertainty of multiple weather measurements across the predefined 38 delivery stations in the small network.}
\label{actual-predicted}
\end{figure*}

\begin{figure*}[!ht]
\centering
\includegraphics[width=0.9\textwidth]{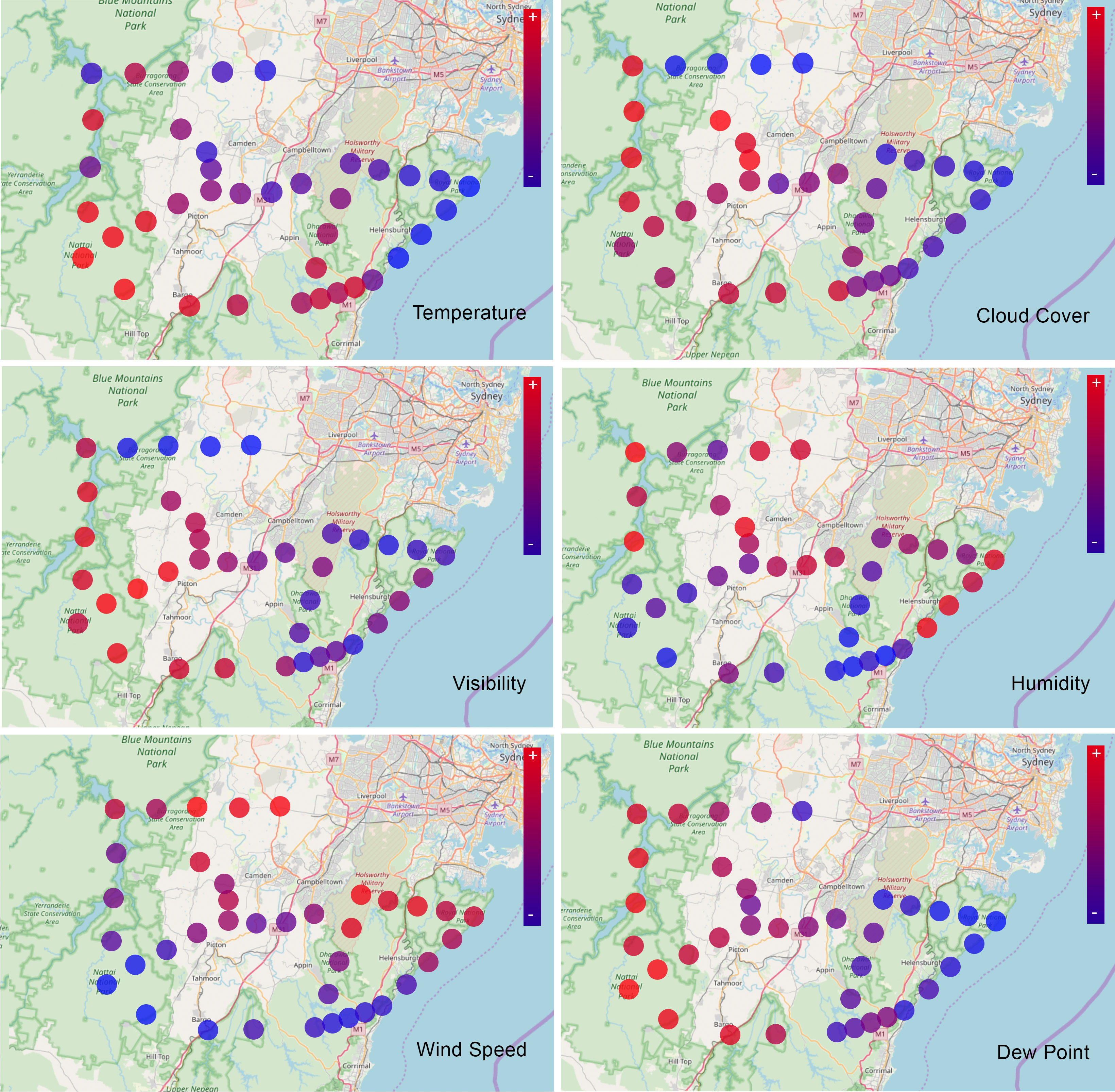}
\caption{Geo visualisation for the uncertainty increases and decreases for each weather factor at the 38 delivery stations. Cloud cover and humidity have fewer measurement deviations. Visibility has large positive deviation shown in red-coloured circles. Dew point got more negative deviations as shown with the blue colour.}
\label{actual-predicted-maps}
\end{figure*}

Fig. \ref{actual-predicted}, shows the average of the differences between the actual and forecast weather measurements for each delivery station. Fig. \ref{actual-predicted-maps} projects increases and decreases on the delivery map. The increasing and decreasing numbers are represented by a colour-gradient between red and blue, respectively. The circles are positioned on the location of each delivery station. 
{The figures show that the visibility forecast measurements are always higher than the actual measurement. The wind speed is also positively increasing in all cases. The temperature tends to be raised at most stations. To assure a high level of certainty of such cases, we subtract the \emph{CM} to the forecast measurements. On the contrast, the forecast of dew point is lower than the actual measurement in most stations. Therefore, we add \emph{CM} to the dew point forecast measurement. The humidity and cloud cover slightly go up and down around the zero. Thus, we use their measurements without any process.}

%

\section{DaaS Dataset}\label{Appendix_data}
In this Appendix, we show more details about the generated DaaS dataset. We generate three main datasets for large and small scale network. 
\begin{itemize}
    \item First, DaaS Skyways network has $2,560$ and $128$ segment-connections between the network nodes for the large and small networks, respectively. Table \ref{network-Skyways} shows examples of the constructed Skyways.
    \item Second, DaaS services have $16,770$ and $30,476$ complete-paths that consist of several movements among the network station, for both the large and small networks, respectively. Table \ref{DaaS-service} lists samples of the DaaS services. These services refer to the virtual drones available at certain locations in particular timestamps. 
    \item Third, the later selections are in total of $4,589,290$ and $478,398$ DaaSs for the large and small delivery networks, respectively.
\end{itemize}

{
\begin{table}
\centering
\caption{Examples of the constructed DaaS network Skyways.}
\label{network-Skyways}
\begin{tabular}{|p{0.5cm}|l|p{1cm}|p{2cm}|p{2cm}|p{2.5cm}|}
\hline
ID & Skyway       & Source & Desti-nation & Distance (km) & Compass bearing \\ \hline
1  & F-DS\_30      & F      & DS\_30      & 8.19     & 6                \\ \hline
2  & DS\_30-DS\_31 & DS\_30 & DS\_31      & 8.03     & 2                \\ \hline
3  & DS\_32-DS\_33 & DS\_32 & DS\_33      & 11.20    & 42               \\ \hline
4  & DS\_13-DS\_16 & DS\_13 & DS\_16      & 10.57    & 170              \\ \hline
5  & A-DS\_20      & A      & DS\_20      & 18.92    & 224              \\ \hline
6  & DS\_27-DS\_29 & DS\_27 & DS\_29      & 14.21    & 242              \\ \hline
7  & DS\_32-DS\_38 & DS\_32 & DS\_38      & 16.21    & 96               \\ \hline
8  & DS\_24-DS\_25 & DS\_24 & DS\_25      & 9.93     & 268              \\ \hline
9  & DS\_15-DS\_16 & DS\_15 & DS\_16      & 3.69     & 71               \\ \hline
10 & DS\_11-DS\_12 & DS\_11 & DS\_12      & 9.05     & 233              \\ \hline
\end{tabular}
\end{table}}

{
\begin{table*}[h!]
\small
\centering
\caption{A sample of the generated DaaS delivery service dataset.}
\label{DaaS-service}
\begin{tabular}{|l|l|l|p{1cm}|p{1cm}|p{1cm}|p{1.5cm}|p{1.2cm}|p{1cm}|p{5cm}|}
\hline
ID & Drone & Source & Desti. & Total distance & Flying Duration & Main-tenance time & Start time & Skyway count & stations \\ \hline
1 & M200 & E & A & 169.19 km & 122.6 min.& 75 min. & 11/1/17 1:00 & 16 & E-DS\_36-DS\_37-DS\_38-DS\_32-DS\_31-DS\_27-DS\_29-DS\_26-DS\_25-DS\_24-DS\_13-DS\_16-DS\_17-C-DS\_20-A \\ \hline
2 & M200 & E & A & 158.31 km & 114.7 min.& 70 min. & 11/1/17 1:00 & 15 & E-DS\_28-DS\_30-DS\_29-F-DS\_26-DS\_25-D-DS\_24-DS\_13-DS\_12-B-DS\_11-DS\_10-DS\_9-A \\ \hline
\end{tabular}
\end{table*}
}

{
\begin{table}[h!]
\centering
\caption{DaaS delivery services movements.}
\label{DaaS-simulation}
\begin{tabular}{|p{0.5cm}|p{2cm}|p{2.5cm}|p{1.5cm}|p{2.4cm}|p{2.2cm}|p{2cm}|}
\hline
ID & Flying duration & \makecell{Arrival\\ time} & Sou-rce & Desti-nation & Skyway number & Total Skyways \\ \hline
1 & 2.7 & 2017-11-01 01:02:45 & E & DS\_36 & 1 & 16 \\ \hline
1 & 2.2 & 2017-11-01 01:20:01 & DS\_36 & DS\_37 & 2 & 16 \\ \hline
1 & 4.6 & 2017-11-01 01:39:37 & DS\_37 & DS\_38 & 3 & 16 \\ \hline
1 & 11.7 & 2017-11-01 02:06:22 & DS\_38 & DS\_32 & 4 & 16 \\ \hline
1 & 6.0 & 2017-11-01 02:27:22 & DS\_32 & DS\_31 & 5 & 16 \\ \hline
1 & 13.2 & 2017-11-01 02:55:34 & DS\_31 & DS\_27 & 6 & 16 \\ \hline
1 & 10.2 & 2017-11-01 03:20:52 & DS\_27 & DS\_29 & 7 & 16 \\ \hline
\end{tabular}
\end{table}
}

\begin{figure*}[!ht]
\centering
\includegraphics[width=0.9\textwidth]{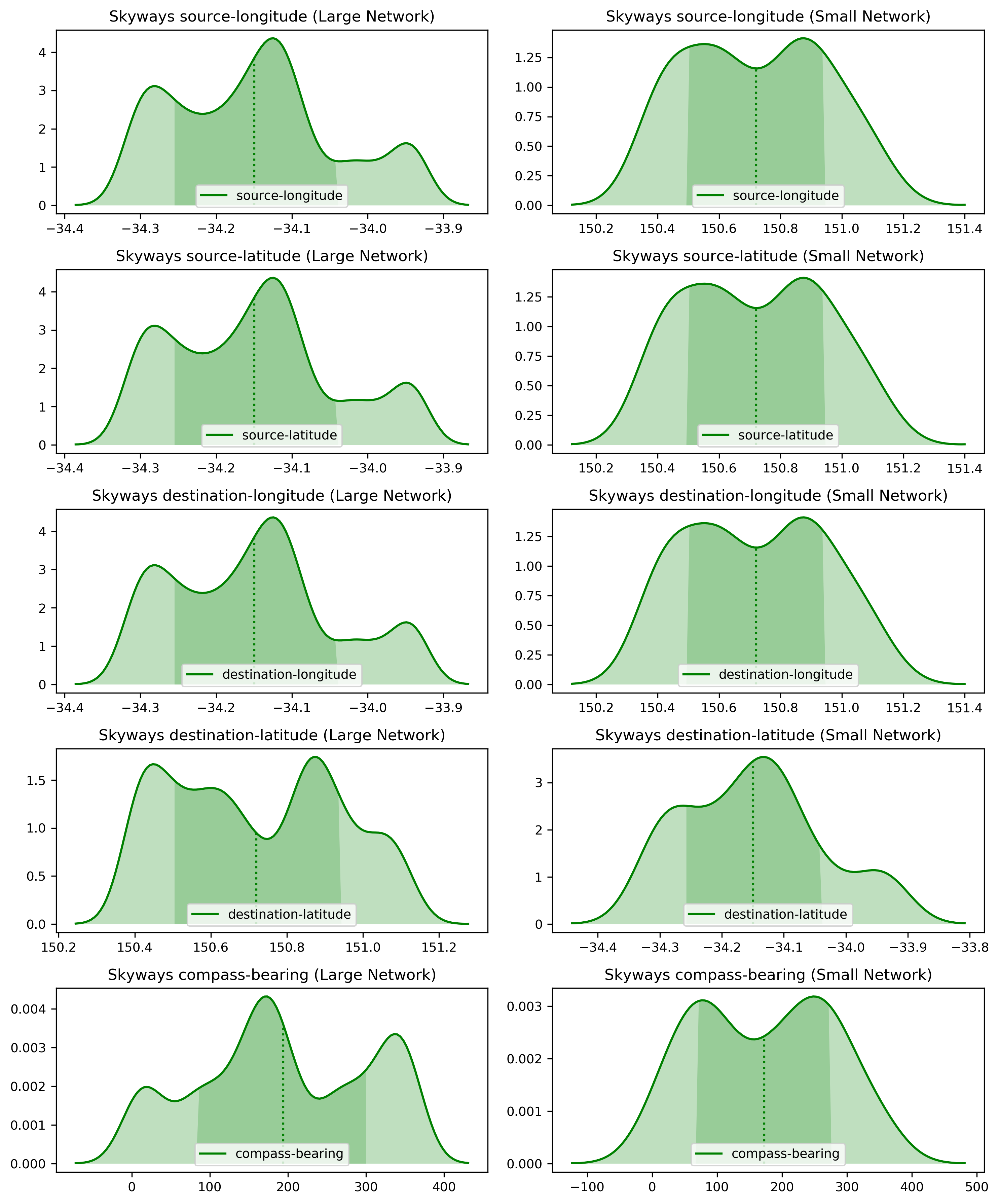}
\caption{DaaS Skyways (segments) has different distributions in terms of different attributes such as geo-location longitude and latitude of source and destination stations, and compass-bearing direction of the segment.}
\label{Seg_dist}
\end{figure*}

Fig. \ref{Seg_dist} shows DaaS Skyways (segments) has different distributions in terms of different attributes such as geo-location longitude and latitude of source and destination stations, and compass-bearing direction of the segment. Both the large and small network have normal (symmetrical) distributions are symmetrically or equally distributed across the mean. However, the two networks have different distribution-modes. The small network mostly has bi-modals uni-from distribution, and the large one has multi-modal ones. These Multi-modal distributions are combining multiple distribution modes around different geo-locations.

\begin{figure*}[!ht]
\centering
\includegraphics[width=0.9\textwidth]{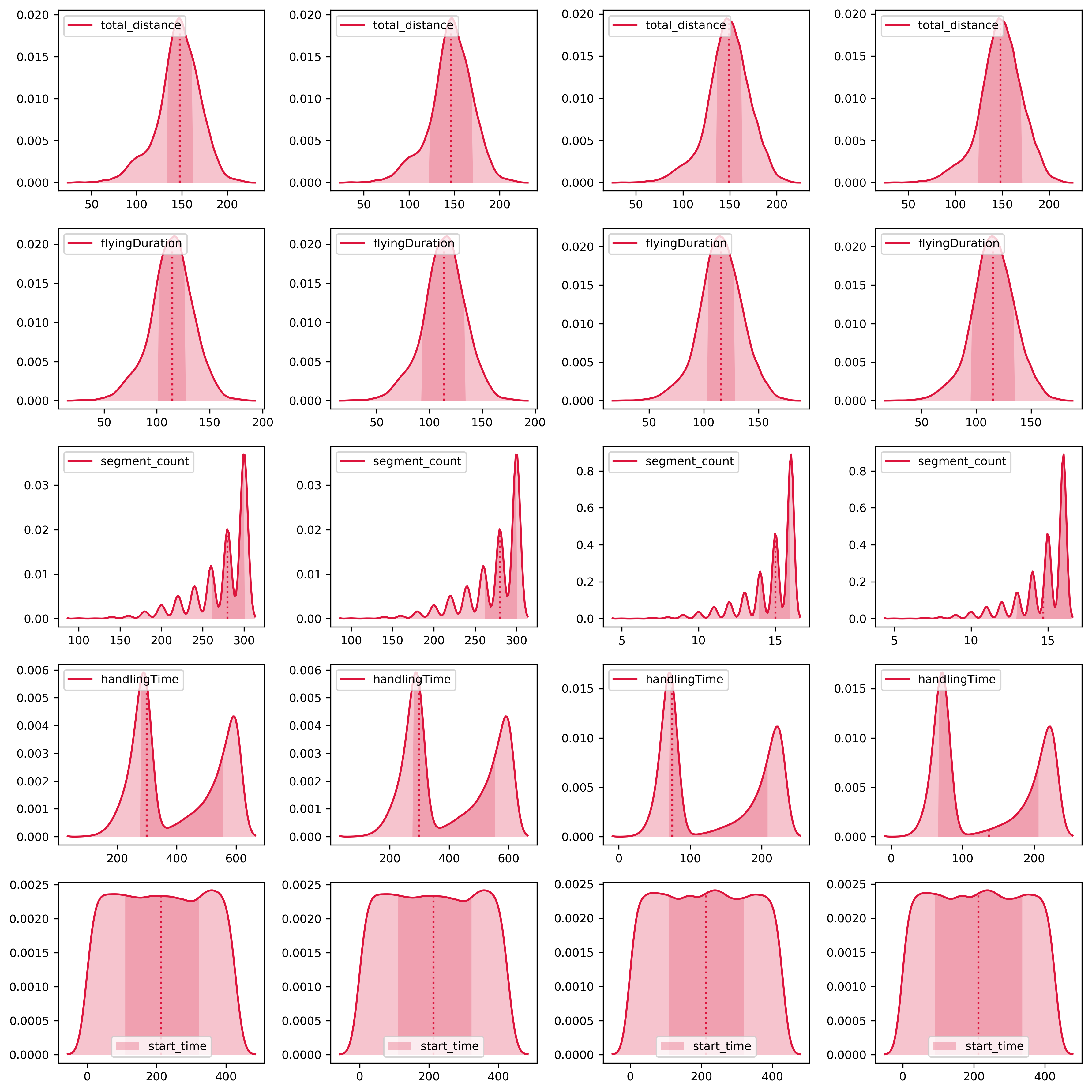}
\caption{DaaS services' distributions of different attributes such as distance, flying duration, handling time, and DaaS start time. The Figure has four columns of graphs visualising (from left to right): 
1) distribution with mean and standard deviation on the large network,
2) distribution with data percentile on the large network,
3) distribution with mean and standard deviation on the small network, and
4) distribution with data percentile on the small network.}
\label{Ser_dis}
\end{figure*}

Fig \ref{Ser_dis} shows DaaS services' distributions of different attributes such as distance, flying duration, handling time, and DaaS start time. The Figure has four columns of graphs visualising (from left to right): 
1) distribution with mean and standard deviation on the large network,
2) distribution with data percentile on the large network,
3) distribution with mean and standard deviation on the small network, and
4) distribution with data percentile on the small network.
Generally, the graphs are not perfectly symmetric, but they show uniform distribution around the mean.
In terms of segment count, the distribution graphs are left-skewed (negative-skewed) having bells at high counts (right-side) with tails at low segment counts. This insight shows that high counts of selection are frequently found in the composite DaaS.

\begin{figure*}[!ht]
\centering
\includegraphics[width=0.9\textwidth]{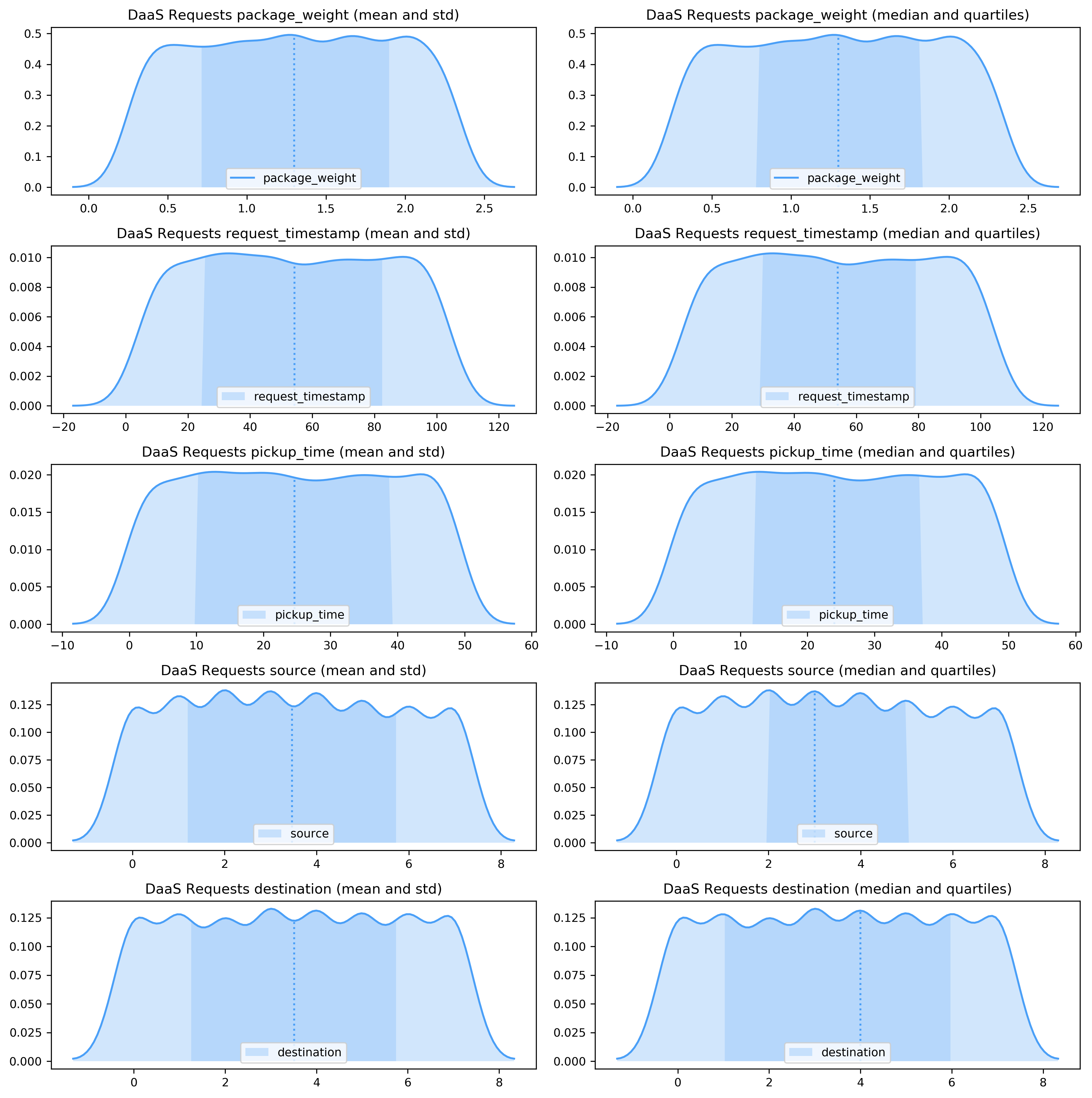}
\caption{DaaS delivery requests show uni-from distribution over all the data attributes.}
\label{PDR_dist}
\end{figure*}

Fig. \ref{PDR_dist} highlights the DaaS delivery requests distributions. The delivery requests generated data show uni-from distribution over all the data attributes. The DaaS requests are symmetrically distributed over delivery weight, request time, geo-locations of source and destination, and pick-up time.


\end{appendices}


\end{document}